\documentclass[preprint,12pt]{elsarticle}

\makeatletter
\def\ps@pprintTitle{%
  \let\@oddhead\@empty
  \let\@evenhead\@empty
  \def\@oddfoot{\reset@font\hfil\thepage\hfil}
  \let\@evenfoot\@oddfoot
}
\makeatother

\usepackage{amssymb}
\usepackage{amsmath}
\usepackage{kantlipsum}
\usepackage{bm}
\usepackage{fancyvrb}
\usepackage{cprotect}

\usepackage{hyperref}
\usepackage{algorithm}
\usepackage[noend]{algpseudocode}

\usepackage{txfonts}
\AtBeginDocument{\mathcode`v=\varv} 

\biboptions{numbers,sort&compress}

\usepackage[normalem]{ulem}

\usepackage{appendix}

\makeatletter
\def\BState{\State\hskip-\ALG@thistlm}
\makeatother

\newcounter{bla}

\usepackage{url}
\usepackage[normalem]{ulem}\usepackage[normalem]{ulem}
\usepackage[usenames,dvipsnames]{color}
\usepackage[capitalize]{cleveref}

\newcommand{\editor}[2]{%
  \expandafter\newcommand\csname #1note\endcsname[1]{%
    \textcolor{#2}{(\textbf{#1:} ##1)}}%
  \expandafter\newcommand\csname #1\endcsname[1]{%
    \textcolor{#2}{##1}}%
  \expandafter\newcommand\csname #1cancel\endcsname[1]{%
    \textcolor{#2}{\sout{##1}}}%
  \expandafter\newcommand\csname #1change\endcsname[2]{%
    \textcolor{#2}{\sout{##1} ##2}}%
  \newenvironment{#1text}{\color{#2}}{\color{black}}
}

\def\qe{\textsc{Quantum ESPRESSO}}
\def\qeheat{\texttt{QEHeat}}
\def\exampleurl{https://gitlab.com/QEF/q-e/-/tree/master/QEHeat/examples}
\def\docurl{https://gitlab.com/QEF/q-e/-/tree/master/QEHeat/Doc}
\def\erf{\mathrm{erf}}
\def\erfc{\mathrm{erfc}}

\definecolor{tangerine}{rgb}{0.944,0.522,0}
\definecolor{greenA}{rgb}{0.,0.844,0.5}

\editor{C}{red}

\usepackage{subfiles}
\usepackage{mathtools}

\usepackage{multirow}

\usepackage{lineno}
\begin{document}

\newcommand{\ai}{\textit{ab initio} }
\def \conduction {\text{\,c}}
\def \valence {\text{\,v}}
\def \barphibm {\bar{\bm \phi}_v^\conduction}
\def \barphi {\bar{\phi}_v^\conduction}

\begin{frontmatter}



\title{\qeheat: An open-source energy flux calculator for the computation of heat-transport coefficients from first principles}


\author[a]{Aris Marcolongo}
\author[b]{Riccardo Bertossa}
\author[b]{Davide Tisi}
\author[b,d]{Stefano Baroni\corref{author}}

\cortext[author] {Corresponding author:\\\textit{E-mail address:} aris.marcolongo@gmail.com (A. Marcolongo) \\\textit{E-mail address:} baroni@sissa.it (S. Baroni) }
\address[a]{Mathematical Institute, University of Bern}
\address[b]{SISSA – Scuola Internazionale Superiore di Studi Avanzati, Via Bonomea 265, 34136 Trieste, Italy}
\address[d]{CNR – Istituto Officina dei Materiali, SISSA, 34136 Trieste}

\begin{abstract}
We give a detailed presentation of the theory and numerical implementation of an expression for the adiabatic energy flux in extended systems, derived from density-functional theory. This expression can be used to estimate the heat conductivity from equilibrium \emph{ab initio} molecular dynamics, using the Green-Kubo linear response theory of transport coefficients. Our expression is implemented in an open-source component of the \qe\ suite of computer codes for quantum mechanical materials modeling, which is being made publicly available.
\end{abstract}


\end{frontmatter}



\noindent {\bf PROGRAM SUMMARY}

\smallskip
\begin{small}
\noindent {\em Program Title:}  \qeheat   \\
{\em Licensing provisions:}  GPLv3 \\
{\em Programming language:}  Fortran \\
{\em Nature of problem:}  The computation of thermal transport coefficients via equilibrium molecular dynamics and the Green-Kubo theory of linear response requires the definition of a heat-flux describing the instantaneous flow of energy. When considering predictive  first-principles methods, a definition of the heat-flux compatible with density-functional theory is required [1]. The evaluation of such a heat flux requires an extension of state-of-the-art atomic simulation codes. \\
{\em Solution method:}  This work describes in detail the numerical implementation of the adiabatic energy current derived in Refs. [1,2] and makes it available to the users of the \qe\ suite of computer codes [3]. Used in conjunction with the \texttt{cp.x} code, to perform Car-Parrinello \emph{ab initio molecular dynamics}, and the \texttt{SPORTRAN} post-processing tool for data analysis [4-6], the program allows to estimate heat transport coefficients in extended systems entirely from first principles. The new code provides as well to developers a modular and easily extendable framework to evaluate time derivatives of electronic properties (e.g. electronic densities or potentials) via a finite difference approach.

\end{small}

\section{Introduction}\label{sec:intro}
The Green-Kubo (GK) linear-response theory of heat transport \cite{Green1952,Green1954,Kubo1957a,Kubo1957b} has long been deemed incompatible with modern quantum-simulation methods  based on Density-Functional Theory (DFT) \cite{Baroni2020,Grasselli2021}, essentially because the concepts of energy density and current, upon which that theory stands, are ill-defined at the molecular scale. This predicament has been reversed by a recent paper \cite{Marcolongo2016} by Marcolongo, Umari, and Baroni (MUB) where it was shown that, due to a general \emph{gauge invariance} principle of transport coefficients \cite{Ercole2016}, the heat conductivity is largely independent of the details of the microscopic definition of the energy densities and currents from which it is derived. Leveraging this remarkable finding, an explicit expression for the energy flux, based on DFT, was given in that same paper. This expression was implemented in a private branch of the \qe\ project \cite{QE1,QE2,QE3} and, due to its considerable complexity, it has not been publicly available so far to the scientific community at large. The purpose of the present paper is to provide a detailed derivation of the MUB expression for the DFT energy flux, a description of its implementation in \qe, and to release and document an open-source code for its computation, named \qeheat. \qeheat\ can be easily interfaced to read a dynamical trajectory generated with a code of choice and compute the MUB flux for the corresponding steps. \qeheat\ is already delivered with a user-friendly interface for the \texttt{cp.x} program of the \qe\, distribution, which is routinely used to perform Car-Parrinello molecular dynamics simulations \cite{QE1,QE2,QE3}. The combination of \qeheat, \texttt{cp.x} and the post-processing tool \texttt{SPORTRAN} \cite{SporTran}, designed to perform the statistical analysis needed to evaluate transport coefficients, provides a convenient framework to compute the heat conductivity of extended insulating systems---be they crystalline, amorphous, or liquid---entirely from first principles. We warn the user that the MUB flux is an \emph{adiabatic} energy flux. This refers to the fact that electrons are supposed to populate the ground state during time evolution, which is the case for insulators with a finite band gap.\\
\\
We note that other computational efforts have been proposed to evaluate microscopic expressions for the energy flux based on DFT. For example, Kang et. al 
\cite{Kang2017} evaluate an atomic decomposition of the total energy and then use a two-step procedure to take into account periodic boundary conditions when evaluating the energy flux. Carbogno et. al \cite{Carbogno2017} neglect the convective contribution to the energy flux and use a DFT-based expression for the virial component. This approach is developed as an approximation suited to the description of thermal transport in solids. The open-source distribution of \qeheat\ will make it easier to compare the computational advantages of the various expressions which, after the assessment of the principle of gauge invariance, are being developed by the community. Finally, the evaluation of the energy flux provided by \qeheat\ involves a finite difference evaluation of electronic properties, e.g. electronic densities or potentials, which are performed entirely in-memory, storing the results of different total energy calculations into ad-hoc data structures. Therefore, \qeheat\  provides as well to developers a modular and easily extendable framework to evaluate time derivatives of electronic properties via a finite difference approach.

\section {Overview} \label{sec:Theory}
According to the GK theory of linear response \cite{Green1952,Green1954,Kubo1957a,Kubo1957b}, the heat conductivity of an isotropic system is given by:
\begin{equation}
    \kappa =  \frac{1}{3 \Omega k_{B}T^2}\int_{0}^{\infty}\Bigl\langle {\bm{J}}(t)\cdot{\bm{ J}}(0)\Bigr\rangle dt,\label{eq:GK}
\end{equation}
where $\Omega$ is the system volume, $\bm{J}(t)=\int_\Omega \bm j(\bm r,t)d\bm r$, \emph{i.e.} the volume integral of the energy current density, is the energy flux (note that with this definition the flux is extensive), $T$ the temperature,  and $\langle \cdot \rangle$ indicates an equilibrium average over the initial conditions of a molecular trajectory.  \qeheat\ serves the purpose of evaluating $\bm J$ at the DFT level of theory. A formal expression for the energy flux can be obtained by integrating by parts the continuity equation $\dot\epsilon = - \nabla \cdot \bm j$ as \cite{Baroni2020,Grasselli2021}:
\begin{equation}
    \bm{J}=\int_\Omega d \bm{r} \dot \epsilon(\bm r) \bm{r}
 \label{eq:formal},
\end{equation}
where $\epsilon(\bm r)$ is the energy density of the system, whose integral is its total energy. Strictly speaking, Eq. \eqref{eq:formal} is ill-defined in periodic boundary conditions (PBC), which are commonly adopted in molecular simulations. In order to compute it explicitly within PBC, one has to first recast it in a boundary-insensitive form, and $\Omega$ can then be replaced with the volume of the simulation cell in Eqs. (\ref{eq:GK}-\ref{eq:formal}). Once this is done, the DFT energy flux can be cast into the MUB form, called here $\bm{J}^{MUB}$, and discussed in detail in later sections. 

\qeheat\ computes the MUB energy flux as a function of the atomic positions, $\{\bm{R}_{s}\}$ and velocities, $\{\bm{V}_{s}\}$, \emph{i.e.} for any selected snapshot of an \emph{ab initio} molecular dynamics (AIMD) trajectory. 
Despite the complexity of the resulting formula for the energy current, from a practical point of view the use of \qeheat\ relies on a limited number of additional input parameters with respect to a standard \qe\  DFT computation. These are reported in the \verb.energy_current. input namelist, which is shown in figure \ref{fig:energy_current_namelist}. The meaning of all the keywords is explained in more detail in section \cref{sec:UsageBenchmarks}. The only additional parameters are \verb.eta. and \verb.n_max., controlling the Ewald summations, which appear only in classical contributions to the energy current, and \verb.delta_t., a time-discretization parameter, used to perform numerical derivatives. The default values should work for most systems.

The work is organized as follows. In \cref{sec:NumericalImplementation} we show the formulas that \qeheat\ implements for the abiabatic energy flux, and Appendix \ref{sec:NumericalImplementation} is dedicated to the exact numerical schemes. \cref{sec:code_structure} contains an overview of the code, and \cref{sec:UsageBenchmarks} describes the input parameters in more details and shows some stability and implementation checks. Finally, in \cref{sec:calculationExample} we showcase how to use the energy flux time series provided \qeheat\ to evaluate thermal conductivity, using the \texttt{SporTran} data analysis post-processing tool (the data analysis is fully explained in the Jupyter notebook data\_analysis.ipynb in the Supplementary Materials). 
\begin{figure}
    \centering
\begin{Verbatim}[frame=single]
    &energy_current
    delta_t = 1.000,
    file_output = 'current_hz',
    eta = 0.100,
    n_max = 5,
    trajdir = 'traj/cp',
    first_step = 1,
    vel_input_units = 'CP'
 /
\end{Verbatim}
    \caption{Example of the energy\_current namelist. delta\_t is the time used for numerical derivatives. eta and n\_max are the parameters used to converge Ewald sums. trajdir is the prefix of the trajectory files. In this example, the program reads the files "traj/cp.pos" and "traj/cp.vel". first\_step tell the program the first step id to compute. The step ids are part of the trajectory file format. After this namelist the full pw.x input is required. A full example of the input and the documentation of the keywords can be found at \url{\docurl}. See also \cref{sec:UsageBenchmarks}.}
    \label{fig:energy_current_namelist}
\end{figure}

\section{DFT energy flux}\label{sec:NumericalImplementation}
In this section we recall the expression of MUB energy flux \cite{Marcolongo2016} and some of the notation used throughout the paper. For a more extensive and detailed study of the implementations of the many components of the MUB current the reader is referred to \ref{sec:ImplCurr}. $\bm{J}^{MUB}$ is expressed as a sum of five components:
\begin{align}
  \bm{J}^{MUB} & =\bm{J}^{KS} + \bm{J}^{0}+ \bm{J}^{n} + \bm{J}^{H} + \bm{J}^{XC}, \label{eq:J_GK0}
\end{align}
where
\begin{align}
    \bm{J}^{KS} & =\sum_{v} \left (\langle\varphi_{v}| \bm{\hat r}\hat{H}^{KS}| \dot{\varphi}_{v}\rangle + \varepsilon_v \langle\dot{\varphi}_{v}| \bm{\hat r} | \varphi_{v}\rangle \right), \label{eq:J_KS}  \\
    \bm{J}^{0}  & =\sum_{s \bm L} \sum_v \left\langle \varphi_{v} \left|(\bm{\hat r}-{\bm R}_s - {\bm L}) \left(\bm{V}_{s} \cdot \nabla_{s \bm L} \hat{v}^0  \right)\right|\varphi_{v} \right\rangle ,\label{eq:J_0} \\
        \bm{J}^{n} & = \sum_{s}\left[\bm{V}_s  e^0_s +  \sum_{t \neq s}\sum_{L}(\bm{R}_s-\bm{R}_t -\bm{L})\left(\bm{V}_{t} \cdot \nabla_{t \bm L}w_s\right) -\sum_{L\neq 0} \bm{L}\left(\bm{V}_{s} \cdot \nabla_{s \bm L }w_s \right)  \right] \label{eq:J_n} \\
    \bm{J}^{H}  & =\frac{1}{4\pi e^2} \int\dot{v}^{H}(\bm{r}) \nabla v^{H}(\bm{r}) d\bm{r},\label{eq:J_H} \\ \bm{J}^{XC} &=  \begin{dcases} 
        0 & \text{(LDA)} \\ -\int n(\bm{r})\dot{n}(\bm{r}) \bm \partial\epsilon^{GGA} (\bm{r}) d\bm{r} & \text{(GGA)}. \label{eq:J_XC} 
    \end{dcases}
\end{align}
In the following, these components are referred to as the \emph{Kohn-Sham}, \emph{Zero}, \emph{Ionic}, \emph{Hartree} and \emph{Exchange-Correlation} fluxes, respectively. For the insulating systems of interest in this work, the ionic degrees of freedom completely define the state of the system and the electrons populate the ground state, according to the adiabatic approximation. Each time derivative, indicated with the usual \emph{dot} operator, has then to be understood from the implicit dependence on the atomic positions.
We note that a complete understanding of the different components of the MUB energy flux is not needed to perform, as a user, a thermal-conductivity calculation. Here and in the following, we indicate with $\bm L $ the lattice vector. We stress that periodic boundary conditions (PBC) are assumed here over the unit (simulation) cell. This implies that Kohn-Sham orbitals and energies are sampled at the $\Gamma$ point of the Brillouin cell. The notation $\nabla_{s \bm L}$ is a  shorthand for the gradient with respect to displacement of the atom at location $\bm R_s + \bm L$. A summation over $s$ runs over all atoms belonging to the simulation cell. Unless otherwise specified, carets indicate quantum-mechanical operators, as in $\hat{H}^{KS}$ or $\hat{\bm r}$. Following is a brief report of the definition of the most important terms according to their physical meaning. For a more extensive summary of the notation used here and throughout the text the reader is referred to \ref{sec:Notation} (``Notation'').
The ionic energy $e^0_s$ is the sum of the kinetic energy, $\frac{1}{2}M_sV_s^2$, and $w_s$ the classical electrostatic interaction between the $s$-atom and all other atoms in the system. The electronic degrees of freedom are instead described by the instantaneous Kohn-Sham Hamiltonian, $\hat{H}^{KS}$, and its eigenvalues and eigenvectors, the Kohn-Sham energies, $\epsilon_{v}$, and orbitals, $\varphi_{v}$. A summation over $v$ runs over all occupied orbitals. Several quantities, like the electronic number-density distribution, $n(\bm r)$, and Hartree potential, $v^H(\bm r)$, are implicit functions of the wave-functions. $\hat v^0$ is the total  external atomic pseudo-potential, describing the interaction between electrons and nuclei. For a more detailed analysis see \ref{sec:Zero_current}. The symbol $\epsilon^{GGA}$ stands for the generalized gradient approximation (GGA) exchange-correlation local energy per particle and its derivative with respect to density gradients is indicated with $\bm {\partial} \epsilon^{GGA}$, which is a vector whose component along direction $i\in \{x,y,z\}$ is given by $\partial \epsilon^{GGA}(n,\nabla n)/\partial(\nabla_i n)$. In the present version of \qeheat\ only the local density approximation (LDA) and the generalized gradient approximation given by the PBE \cite{PBE} functional are implemented.

The formulas reported are compatible with PBC. Thus, they can be implemented for periodic systems, where some of the summations need to be extended to all the periodic replicas of the atoms. Nevertheless, the computation of the various contributions to the energy current, Eqs. (\ref{eq:J_KS}-\ref{eq:J_XC}), is plagued by the occurrence of several divergences, arising from the long range character of the Coulomb interaction. As it is the case for the total energies, atomic forces, and stress, the individual electronic, ionic, and electron-ion contributions diverge and it is only their sum that is regular in the thermodynamic limit. In order to regularize the individual components of the MUB flux, we compute all the relevant terms by screening the Coulomb interaction with a Yukawa cutoff, $\frac{1}{x} \rightarrow \frac{e^{-\mu x}}{x}$. In \ref{sec:ImplCurr}, we check explicitly that the singular contributions to the various terms cancel each other in the $\mu\to 0$ limit, so that they can be consistently and safely neglected and do not appear in the final formulas reported in \ref{sec:ImplCurr}.

\section{Code structure}\label{sec:code_structure}

We start by describing the strategy implemented to compute numerical derivatives of quantities appearing in Eqs.~\eqref{eq:J_KS}, \eqref{eq:J_H}, and~\eqref{eq:J_XC}, like $\dot n$ and $\dot v^H$, since they require a special treatment. Quite generally, one needs to evaluate terms of the type $\dot f(\{ \bm  R_s(t) \})$, where the function $f$ can be a scalar function, which depends on time only through the set of the instantaneous ionic positions $\{ \bm  R_s(t) \}$, evolving according to Hamilton's equations of motion. \qeheat\ implements a finite-difference scheme, using by default a symmetric numerical differentiation formula:
\begin{equation}
    \dot f(\{ \bm  R_s\}) \approx
        \frac{f(\{\bm  R_s +\bm V_s dt/2\})-f(\{\bm  R_s-\bm V_s dt/2\})}{dt}
      \label{eq:generic}
\end{equation}
The small parameter $dt$ is an input of the computation. In such a scheme quantities that are not differentiated are evaluated at time $t$, so three wave-functions are required to be kept in memory at the same time. \qeheat\ performs therefore for each step two additional self-consistent-field (SCF) DFT calculations, using the same DFT solver of the \qe\ distribution, at slightly displaced positions, i.e. $\{ \bm R_s-\bm  V_s dt/2 \}$ and $\{ \bm R_s+\bm  V_s dt/2 \}$ along the AIMD trajectory.
The wave-functions of the previous calculation are used as a starting point for the next one, which require much less iterations to converge.
We note that \qeheat\ gives the user also the possibility to use a non-symmetric differentiation scheme, which is shown in \ref{sec:dt_ren_2p}. This scheme is computationally cheaper. Nevertheless, for differentiable functions, the order of convergence of the symmetric scheme is quadratic in $dt$, whereas the non-symmetric one is linear. Accordingly, the stability is improved with the default symmetric scheme. We recommend therefore the latter and use it for all calculations here presented. See also the dedicated \ref{sec:dt_ren_2p}. \\

The trajectory data is managed by the Fortran derived data type \verb.cpv_trajectory. defined in the file \verb,cpv_traj.f90,, while
the orbitals and the associated atomic position are managed by the derived type \verb,scf_result,, implemented in \verb,scf_result.f90,. The most relevant subroutine that acts on this object is \verb.scf_result_set_from_global_variable., that copies the eigenfunctions, the eigenvalues, the potential and the atomic positions from the \qe's global variables to the instance of \verb.scf_result.. The results for each of the three (or two) wave-functions that are required by the computation routines are stored in the variable \verb.scf_all., defined in the main program routine. Global variables are avoided as mush as possible.

The code starts by reading the input ``namelists'': first the \verb.energy_current. namelist, then all the \texttt{pw.x} namelists. Then it calls all the \texttt{pw.x}-related initialization routines. After eventually reading the previously generated output file that allows the program to set the correct starting timestep, it enters the main loop over the input trajectory timesteps. The trajectory files have the same format of \qe's \verb_cp.x_ code output files.

The most important routines where the above mentioned data structures are used are the following: 
\begin{itemize}
    \item \verb.SUBROUTINE current_zero. (module \verb.zero_mod.)\\
    Carries out the computation of \cref{eq:J_0}. This routine is called in the middle of the computation using the same timestep $t$ of the positions stored in the input trajectory, so that the result does not depend on $dt$.
    \item \verb.SUBROUTINE current_ionic. (module \verb.ionic_mod.)\\
    Computes all parts of \cref{eq:J_n}, and it is called as \verb.current_zero. at the same timestep of the input trajectory
    \item \verb.SUBROUTINE current_hartree_xc. (module \verb.hartree_xc_mod.)\\
    Computes \eqref{eq:J_H} and \eqref{eq:J_XC}. Since a numerical derivative is needed, this routine reads the wave-functions from the global type \verb.scf_all. and it is run at the end of all necessary \verb.run_pwscf. calls.
    \item \verb.SUBROUTINE current_kohn_sham. (module \verb.kohn_sham_mod.)\\
    Computes \eqref{eq:J_KS}. As \verb.current_hartree_xc., it needs all the wave-functions calculated by the DFT solver for this step.
    \item \verb.SUBROUTINE run_pwscf.\\
    Uses \qe's routines to solve the DFT problem for the atomic positions stored in the global array \verb.tau.. Equivalent (but the starting wave-function and potential, that can be the last computed one) to a standard call to the \verb,pw.x, program with the input stripped of the \verb.ENERGY_CURRENT. namelist. The result is stored in the Quantum Espresso's global arrays (\verb.evc.)
    \item \verb.SUBROUTINE prepare_next_step.\\
    This routine is used to change the global array \verb.tau. to \verb.tau. + \verb.vel.$\cdot dt\cdot$\verb.ipm., where \verb.ipm. is the argument of the subroutine that can be -1,0,1. After doing that it calls the necessary routines to prepare the potential for \verb.run_pwscf..
\end{itemize}
The 4 modules, one for each part of the MUB current, are completely independent of one another.
The structure of the main loop over the trajectory's time steps is summarized in \cref{alg:general_structure}.
\begin{algorithm}
\caption{Workflow of \texttt{all\_currents.f90.} } \label{alg:general_structure}
\begin{algorithmic}[1]
\State \qe\ initialization (plane waves, pseudo-potentials,...) 
\State Reading of Restart
\For{each snapshot}
\State call \verb.run_pwscf. with positions displaced at $t-dt/2$
\State call \verb.run_pwscf. with non-displaced positions at $t$
\State call \verb.current_zero., evaluate currents derived from the pseudo-potential 
\State call \verb.current_ionic., evaluate the electrostatic and kinetic Ionic current 
\State call \verb.run_pwscf. with positions displaced at $t+dt/2$ 
\State call \verb.current_hartree_xc., evaluate Excange and Hartree currents 
\State call \verb.current_kohn_sham., evaluate Kohn-Sham current 
\EndFor
\end{algorithmic}
{\small Steps 6 and 7 do not require any finite differences, while steps 9 and 10 do. Step 10 is the most expensive.}
\end{algorithm}

As every big computational code an extended test suite is needed to safeguard the correctness of the calculation after every source code modification. We implemented small tests that are able to run on a single core of a cheap laptop that check against changes in the numerical output of many parts of the code, using the standard \qe's test suite framework.

To conclude the section we want to do some remarks on the code and its interactivity with others typical \ai simulations tools. In principle the wave-functions computed on-the fly by \texttt{cp.x} during the AIMD run could be used, but we preferred to implement a workflow where the computation of the currents is completely decoupled from the AIMD engine, thus the wave-functions are always recomputed by \texttt{pw.x}. The chosen approach allows the user to run the calculation in post-processing mode, thus using the preferred code to generate the dynamics, not to be limited to those in the QE packages. This way it allows, also, a trivial and powerful per-snaphot parallelization.

\section{Code usage and benchmarks} \label{sec:UsageBenchmarks}
\subsection{Input description} \label{sec:inputDescription}

    The input is organized in a traditional fortran namelist input file, similar to the input files of many Quantum Espresso's programs, and an optional trajectory file (that is a file for the atomic velocities and a file for atomic positions) if the user wants to compute the energy current for more than one snapshot with a single run. A full example of the input can be found at \url{\exampleurl}. Before running \qeheat\ it is necessary to obtain velocities and positions from a different code, for a complete description of the units of measure see Table 1. If the Quantum Espresso's \verb,cp.x, program is used for this purpose, its output trajectory files can be recycled as input trajectory files without any modification. The program's mandatory input is organized into an \verb.ENERGY_CURRENT. namelist and all the usual \verb,pw.x, namelists. We remind the user that, up to the present version, only norm conserving pseupotentials and the PBE exchange correlation functional are supported. At the end of the input file the \verb.ATOMIC_VELOCITIES. card is required. In the \verb.IONS. namelists the value \verb.ion_velocities = 'from_input'. is required, since the program must read the atomic velocities to compute the energy current. An extensive input description can be found in the file \verb,INPUT_ALL_CURRENTS.html, in the \verb.Doc. folder of the code repository. Here we remark the most important parameters of the \verb.ENERGY_CURRENT. namelist:
\begin{itemize}
    \item \verb.delta_t.  : time in PW's atomic unit used to compute all the numerical derivatives like the one in Eq.~\eqref{eq:generic};
    \item \verb.trajdir.  : prefix of the cp-formatted trajectory. Optional: if not setted, only the positions and the velocities of the input file are read;
    \item \verb.n_max. : number of periodic images along the directions of each basis cell vector to converge Ewald sums. This fixes the range of $\bf L$ in Eq.~\eqref{eq:J_n} ;
    \item \verb.eta.  : convergence parameter of the Ewald sums needed in the computation of $\bm J^n$, for more details see \ref{sec:ionic_current}.
\end{itemize}
An example of the namelist is provided in fig.~\ref{fig:energy_current_namelist}. An additional output file is written and updated at the end of each step in the folder where the program is run. All the currents are printed in a column format, ready to be analyzed by an external post-processing tool.

As discussed in \ref{sec:dt_ren_2p}, \verb.CONTROL.'s \verb.conv_thr. and \verb.ENERGY_CURRENT.'s \verb.delta_t. have a profound link and influence heavily each other, and despite we think the default value of \verb,delta_t=1.0, is safe enough, they must be carefully tested, veryfing that the standard deviation of the result is low enough.

The standard deviation of the output energy current can be estimated by repeating the same calculation for every step, many different times, setting for each repetition a random starting potential and a random starting wave-function. The input options \verb;re_init_wfc_1 = .true.; together with \verb.n_repeat_every_step = 20., for example, do 20 repetition of every timestep, resetting the starting wave-functions/potential before the first scf calculation. The \verb;pw.x;'s input option \verb;startingwfc = 'random'; is suggested, to obtain a faithful error estimation. If more reinitializations are desired, the options \verb;re_init_wfc_2; and \verb;re_init_wfc_3; can control the randomness of the starting wfc and potential of every of the 3 (or 2) wave-functions needed to perform the numerical derivatives, as explained in section~\ref{sec:code_structure}. Note that when the wave-function is reinitialized from scratch, the computation time raises since more scf cycles are required to reach the target convergence threshold. When \verb.n_repeat_every_step. is greater than 1, an additional column formatted output file with the averages and the standard deviations is produced.

\begin{table}
    \centering
    \begin{tabular}{c|c|c}
         & parameter & units \\
         \hline
         \multirow{4}{*}{INPUT}& dt &  $\tau_{a.u.}$\\
         &eta & $1/a_0^2$ \\
         &velocities & $a_0/\tau_{a.u.}$ (CP units can be specified) \\
         &positions & $a_0$ \\

         \hline

         \multirow{3}{*}{OUTPUT} & energy current & $Ry\cdot a_0/\tau_{a.u.}$ \\
         &electronic density current & $a_0/\tau_{a.u.}$ \\
         &center of mass currents & $a_0/\tau_{a.u.}$  \\

    \end{tabular}
    \caption{Units used for the input and the output, where $Ry=2.1799\cdot 10^{-18}J=13.606 eV,\;\;a_0=5.2918\cdot 10^{-11}m,\;\;\tau_{a.u.}=4.8378 \cdot 10^{-17}s$ are the Rydberg units of energy, the Bohr radius and the time unit in Rydberg atomic units. The program assumes the input velocities to be in Rydberg atomic units, the standard for \texttt{pw.x}, unless specified otherwise with vel\_input\_units='CP' in the energy\_current namelist. In that case it assumes Hartree atomic units, the standard for \texttt{cp.x}. \texttt{cp.x}'s unit of time is $2.4189\cdot 10^{-17}s$. } 
    \label{tab:units}
\end{table}

\subsection{Implementation checks: Finite systems translating at constant speed}\label{sec:translatingSystems}
\begin{figure}
    \centering
    \includegraphics[width=0.8\textwidth]{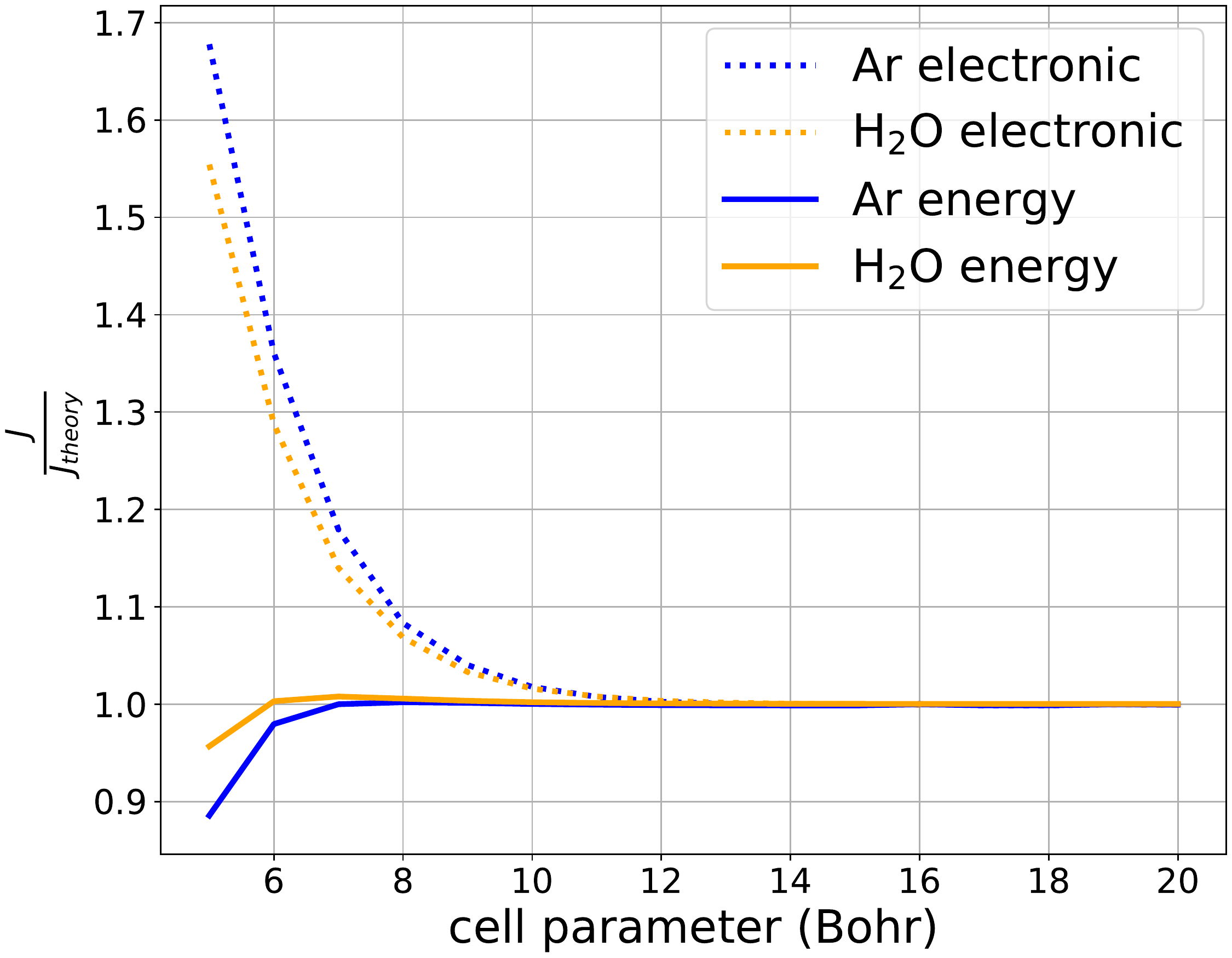}
    \caption{Proof that the behaviour of the code is correct for systems translating at equilibrium and constant velocity. In this setting, the output of QEHeat can be exactly compared with the known theoretical value, indicated with $J_{theory}$, in the large cell limit. Indeed the plot shows that the ratio $J/J_{theory}$ goes to one increasing the simulation cell. This test can be performed for $J$ equal to the electronic density current (dotted), that is used to calculate a part of the energy current, and for the energy current itself (not dotted). Tests are performed for a single relaxed water molecule ($H_2O$) and a single argon atom ($Ar$). The electronic current should be, in the infinite cell limit, $\bm J^{el}_{\text{theory}}=N_{el} \times \bm v$, where $N_{el}$ is the number of electrons and the energy flux should be, in the infinite cell limit, equal to $\bm J^{ene}_{theory}=E^{tot}\times \bm v$, $E^{tot}$ being the total energy.}
    \label{fig:J_translation}
\end{figure}

The Green-Kubo current associated with a localized energy density $\epsilon(\bm r,t)$ rigidly translating with constant velocity $\bm v$, is equal to $E^{tot}\times \bm v$. One possible way to show this is to consider $\epsilon(\bm r,t)=\epsilon(\bm r-\bm v t,0) \equiv \epsilon^0(\bm r-\bm v t)$,thus $J_a=\int \dot \epsilon r_a d \bm r=-v_b \int (\partial_b \epsilon^0) r_a d\bm r=v_a \int \epsilon^0 d\bm r=E^{tot}v_a$. Note that we used the fact that $\epsilon^0$ can be taken identically equal to zero at the boundary of the integration volume, to remove boundary contributions from the integraton by parts. The identity requires therefore the energy density to be localized and this condition can be mimicked in PBC considering a large enough supercell. We used this property to check the correctness of our implementation for each individual current in \cref{eq:J_GK0}. We simulate a single Argon 
atom and a water molecule at equilibrium, both translating at constant speed. We then  compare $\bm J^{MUB}$ output from \qeheat\ and $E^{tot}\bm v$, where $E^{tot}$ 
is evaluated using an independent computation from the QE code. As discussed, the resulting currents need to be equal only in the limit of large cells, where 
boundary effects can be neglected, i.e. the energy density is truly localized, and under tight convergence criteria. In Fig. \ref{fig:J_translation}, we report the ratio between the computed and theoretical values as a function of the cell parameter, showing that the correct limit behavior is recovered. For this calculation we used a cutoff of $120 Ry$ and $\text{econv}=10^{-14} Ry$. Additionally, in Appendix C  we perform the same test using a large cell parameter but removing individual current components (i.e. the \{\text{XC},\text{IONIC},\text{ZERO},\text{KOHN}\} components) from the total energy flux. Since removing each component changes the difference with respect to the limiting theoretical value, this proves the correct implementation of each individual current. 

In the same figure, using the same approach, we tested the electronic density current defined in eq.~\eqref{eq:el-current} as well. In the infinite cell limit the electron density current of a system translating at a constant speed $\bm v$ is $\bm J^{el}=N_e \bm v$ where $N_e$ is the number of electrons. It is possible to see in figure~\ref{fig:J_translation} that the correct limit is obtained, validating the implementation of the code.

\section{Example of a thermal conductivity calculation}\label{sec:calculationExample}

As an example of a complete \qeheat\ calculation, coupled with the \texttt{SPORTRAN} tool for signal analysis, we report the results of a thermal conductivity calculation for PBE (\cite{PBE} ) water. It is known in the literature \cite{Chen10846,GalliWater,MarzariWater} that DFT with PBE functional does not provide a correctly structured water.
Nevertheless, PBE water is a simple example of a single component molecular fluid that can be used to show all aspects of the theory. For a complete description of the signal-analysis theory used here to post-process the simulation, based on cepstral analysis, we refer to the works \cite{Baroni2020,Grasselli2021, Ercole2016, Marcolongo2020} and to \cite{SporTran} for example notebooks and tutorials. The details of the present analysis are reported for reproducibility in the Jupyter notebook "$\text{data}\_\text{analysis.ipynb}$", provided as supplementary material, and we report here the results. 
We warn the reader that the heat flux returned by \qeheat\ almost always contains non-diffusive signals, i.e. signals not contributing to thermal conductivity, with a high amplitude, that may be a problem for a standard numerical Green-Kubo integration. The high formation energies present in ab-initio computations can generate such components in the energy flux. By using the fact that these components are highly correlated with the mass and electronic density flux, renormalization or multicomponent techniques, effectively decorrelating the energy flux from these spurious signals, solve this issue. For a longer discussion we refer to \cite{Marcolongo2020}  and references therein. Here we took advantage of the multicomponent filters implemented in \texttt{SPORTRAN}. \\
We consider a 64 molecules water system at 600K. First,  the molecular dynamics trajectory was computed with the cp.x code using a cubic box with size of 12.43\AA  ~and an integration timestep $\Delta t=3 $ in Hartree atomic units, with a total length of 240ps. 
\qeheat\ shows excellence stability properties as a function of the time-discretization parameter $dt$, which are shown and thoroughly discussed in \ref{sec:dt_ren_2p}.  A value of $dt$ equal to twice the Car-Parinello MD simulation timestep, $dt=2\Delta t$, is of particular interest because it would be beneficial in an on-the-fly computation, allowing to reuse the same wavefunctions computed in the MD simulation, neglecting the need for the recomputation of the scf cycles. In the Appendix we also show that,  at a fixed value of the self consistency threshold for convergence \texttt{econv}, the variance of the MUB estimator decreases with $dt$. Thus, it is favourable, in order to control numerical noise, to choose $dt$ as large as possible, as long as the bias introduced by nonlinear effects is negligible. For these calculations, we first investigated two values of \texttt{econv}, namely $10^{-8}$ and $10^{-11}$ Ry with a rather small $dt=0.5$ Rydberg  atomic units. The final result of the computation, i.e. the thermal conductivity coefficient, is affected very weakly by this parameter, that on the contrary changes in a noticeable way the computational cost. In this case both computations gave $(1.07 \pm 0.09)$W/mK. Moreover we checked that using a $dt$ that is equal to the integration timestep of the Car-Parrinello simulation and $\text{econv}= 10^{-8}$ produces the same spectrum with no noticeable differences and the same thermal conductivity coefficient: in this case we had $(0.98 \pm 0.09)$W/mK.  

\begin{figure}
    \centering
    \includegraphics[width=0.8\textwidth]{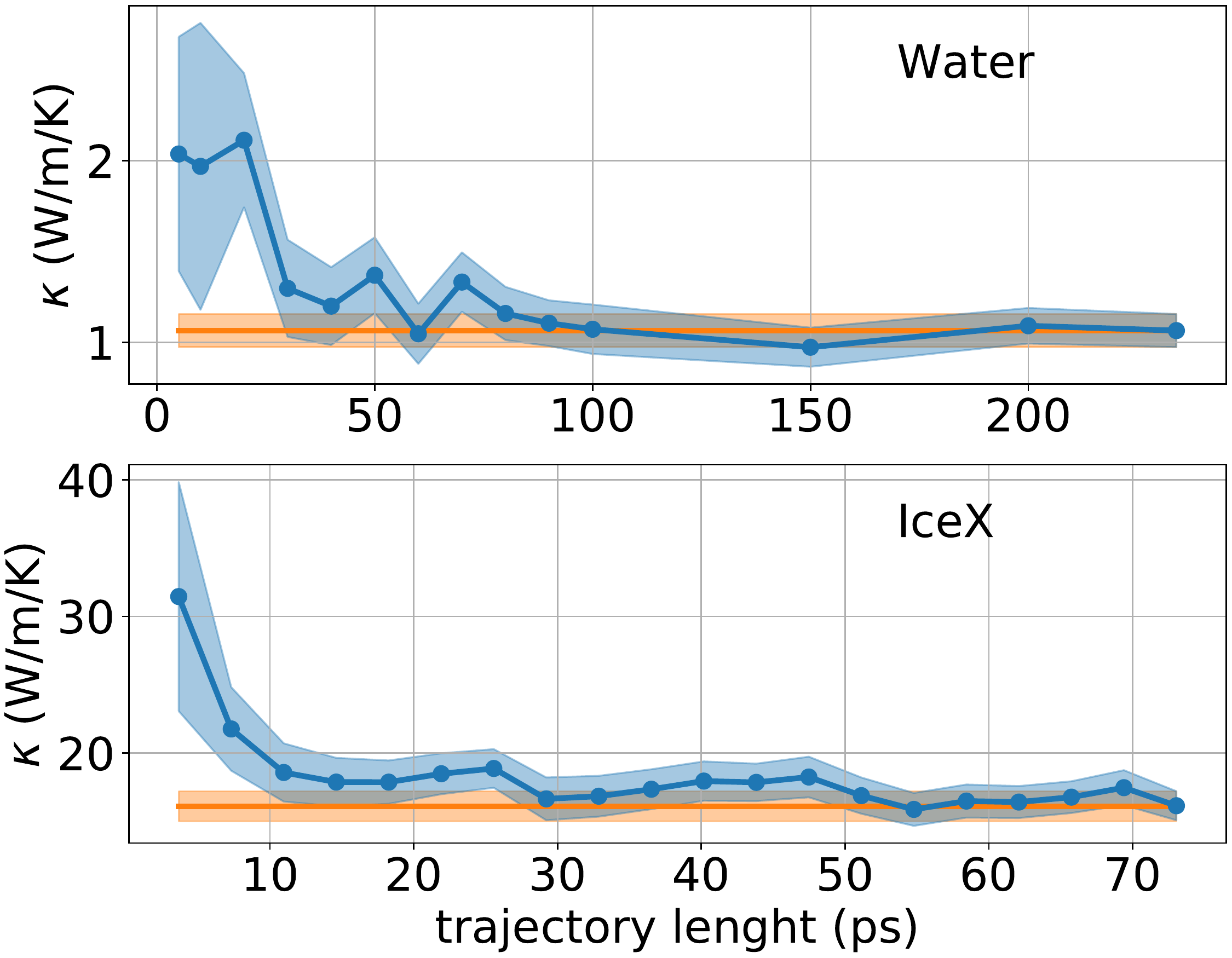}
    \caption{The dependence of $\kappa$ with respect to the trajectory length for 64 molecules of water system at $600$K and for the ice X system in \cite{Grasselli2020}. The orange line correspond to the values with the longest trajectory possible, $\kappa = 1.07 \pm 0.09~$W/(mK) and $\kappa = 16.1 \pm 1.1~$W/(mK) for water and ice X respectively. The data for the ice X system are taken from the Material Cloud repository \cite{GrasselliMaterialCloud}}
    \label{fig:Water_IceX_vs_ps}
\end{figure}

According to our experience, the overall cost of a \qeheat\ calculation is often of the same order of magnitude of the full Car Parrinello molecular dynamics simulation even if the \qeheat\ computation can be trivially parallelized. The data analysis with \texttt{SPORTRAN} adds a negligible computational cost. A more detailed study of the computational costs and how to reduce it can be found in \ref{sec:Comp_Cost}. 
Here we just stress that the total computational time can be also reduced choosing an appropriate length of the simulation, according to the desired precision. \cref{fig:Water_IceX_vs_ps} show the dependence of $\kappa$ from the simulation length both for our water system (upper panel) and a solid ice X structure from ref. \cite{Grasselli2020} (lower panel), where the oxygen are packed in a bbc lattice and tetrahedrally coordinated to hydrogen atoms located exactly midway between two neighboring oxygen atoms. In the present didactic work we choose a very long simulation time of $240~$ps but from the figure is it clear that a simulation length around $100~$ps can provide a reasonable estimate. Depending on the system, even shorter simulations lengths could suffice: the lower panel of \cref{fig:Water_IceX_vs_ps} presents the same data of ice X from \cite{Grasselli2020}, where a simulation of $20-30~$ps proved to be sufficient. We also warn the reader that for strongly harmonic systems, e.g. solids at ambient temperature and pressure, also large cell sizes may be required to remove boundary effects.

\section{Conclusions}
We have presented a robust implementation of the MUB energy current, allowing to compute the heat conductivity from an explicit expression of the energy flux based on DFT, something that is made possible only by recent findings \cite{Marcolongo2014}. The code is distributed via \qeheat, an open-source component of the \qe\ suite of computers codes, and it interfaces easily with other major components of  the distribution, mainly \verb|pw.x| and \verb|cp.x|. The implementation has been tested and validated, proving it to be stable and solid with the parameters. Finally, \qeheat\ can also be considered as a modular and easily extendable informatic framework to evaluate time derivatives of electronic properties, e.g. electronic densities or potentials.

\section*{Declaration of competing interest}
The authors declare that they have no known competing financial interests or personal relationships that could have appeared to influence the work reported in this paper.

\section*{Acknowledgments}
 The authors are grateful to Federico Grasselli, Paolo Pegolo and Pietro Delugas for a thorough reading of the manuscript, and to Pietro Delugas and Loris Ercole for early contributions to the code base. This work was partially funded by the EU through the \textsc{MaX} Centre of Excellence for supercomputing applications (Project No. 824143) and by the Italian Ministry for University and Research (MUR), through the PRIN grant FERMAT.

\bibliographystyle{elsarticle-num}
\bibliography{ShortTitles,currents}







\appendix

\section{Notation}
\label{sec:Notation}
Notations and definitions used throughout the text :
\begin{itemize}
   \setlength\itemsep{0pt}
   \item $e$ : electron charge ;
   \item  $e^0_s$ : ionic energy, equal to $\frac{1}{2}M_{s}v_{s}^{2} + w_{s}$, where $w_s$ is the electrostatic energy.
   \item $\epsilon_{XC}$ : local XC energy per particle, defined by the relation: $E_{XC}= \int \epsilon_{XC} [n](\bm{r})n(\bm{r})d\bm{r}$.  ``LDA'' and
  ``GGA'' in Eq.~\eqref{eq:J_XC} indicate the local-density and generalized-gradient approximations to the XC energy functional ;
   \item  $\epsilon_{v}$ : electronic eigenvalues ;
   \item $\tilde{f}(\bm G)$ : Fourier transform of periodic functions. Given a function $f(\bm r)$ periodic with respect to the unit cell, the Fourier transform $\tilde{f}(\bm G)$, evaluated at the reciprocal lattice vector $\bm G$ is defined by the convention:
   \begin{equation}
    \tilde{f}(\bm G) \equiv \frac{1}{\Omega} \int_{\Omega}  f(\bm r)e^{-i \bm G \cdot \bm r}d \bm r ;
   \end{equation}
   \item $\mathcal{F}[f](\bm  G)$ : Fourier transform of non periodic and localized functions. Given a function $f(\bm r)$ smooth enough and localized (hence not periodic), its Fourier transform is defined for every value of $\bm G$ (hence not only for reciprocal lattice vectors), by the convention:
   \begin{equation}
    \mathcal{F}[f](\bm  G) \equiv \frac{1}{\Omega} \int_{\bm{R^3}}  f(\bm r)e^{-i \bm G \cdot \bm r}d \bm r,
   \end{equation}
   where a convenient normalization factor $\Omega$ has been introduced ;
   \item $\hat{H}^{KS}$ : instantaneous Kohn-Sham (KS) Hamiltonian ;
   \item $\bm L$ : lattice vector, as defined by the unit cell ;
   \item $M_s$ : atomic mass of atomic $s$ ;
   \item $n(\bm{r})$ : ground-state electron-density distribution, defined as $n(\bm r)=\sum_v |\phi_v|^2$ ;
   \item $v^{H}$: Hartree potential, defined as:
   \begin{equation}
    v^{H}(\bm r)=\int_{\bm{R}^3} \frac{e^2n(\bm r')}{|\bm
     r -\bm r'| }d\bm r'.
   \end{equation}
   Note that if $n(\bm r)$ is periodic, also the Hartree potential is periodic;
   \item $\bm{\hat r}$ : multiplicative position operator or versor along direction pointed by vector $\bm{r}$. The meaning should be clear from the context.
     \item $\bm{\hat v}$ : unit versor along the direction of $\bm v$
   \item $\hat v_0$ : ionic (pseudo-) potential acting on the electrons ;
   \item $w_{s}$ :  electrostatic energy, equal  to
   $w_s = \frac{e^2}{2} \sum_{t\neq s}\sum_{\bm L}\frac{Z_tZ_s}{\vert \bm{R}_s-\bm{R}_t-\bm{L}\vert} + \frac{1}{2}e^2 Z_s^2\sum_{ L\neq 0}\frac{1}{L}$ ;
   \item $Z_s$ : atomic charge of atom $s$, expressed in units of the elementary electronic charge ;
   \item $\phi$, $|\phi \rangle$  : a generic normalized wave-function. With brackets, the same wave-function is considered as a vector of a Hilbert space with a scalar product $\langle \: \rangle$ ;
   \item $\Omega$ : volume of the unit cell ;
   \item $\bm \partial\epsilon_{GGA}$ : derivative of the GGA XC local energy per particle with respect to density gradients. It is a vector whose component along direction $i$ is explicitly given by $\partial \epsilon_{GGA}/\partial(\nabla n_i)$ ,
   \item $\nabla$ : gradient with respect to the spatial coordinate $\bm{r}$ ;
   \item $\langle \: \rangle$ : standard scalar product between wave-functions ;
   \item $\dot{[\:]}$ :  derivative with respect to time ;
\end{itemize}

\section{Numerical implementation of each current component}\label{sec:ImplCurr}
In the following appendix we will give an exhaustive and detailed description of the implemented components of the MUB current.
\subsection{Kohn-Sham current}
Starting from Eq.~\eqref{eq:J_KS} and after some simple algebra \cite{Marcolongo2014} we end up with the following expression for the Kohn-Sham current. For every Cartesian component, $i$, one has:
\begin{equation}
{J}^{KS}_i = \mathcal{R} \sum_v \langle \barphi\,_i(t)\left|\hat{H}^{KS}+\epsilon_v\right|\dot \phi _v^\conduction(t)\rangle,
\label{eq:J_KS_exp}
\end{equation}
where
\begin{align}
    | \barphi\,_i \rangle & \equiv \hat{P}^{\conduction} {\hat r}\,_i\left|\phi_v\right>, \\
    |\dot\phi_v^\conduction \rangle,  & \equiv\hat{P}^\conduction\left|\dot\phi_v\right>, \\
    \hat{P}^\conduction &\equiv 1-\hat{P}^\valence \\ \hat{P}^\valence&\equiv \sum_v \left| \phi_v \right> \left<\phi_v\right|,
\end{align}
and $\hat{P}^\valence$ and $\hat{P}^\conduction$ are the projectors over the occupied- (valence-) and empty- (conduction-) state manifolds respectively. $\barphi$ is calculated by solving the linear system:
\begin{equation}
(\hat{H}^{KS}-\epsilon_v + \alpha \hat{P}^\valence)\left|\barphi\,_i\right> = \hat{P}^\conduction[\hat{H}^{KS}, {\hat r}\,_i]\left|\phi_v\right>, \label{eq:barphi_v^c}
\end{equation}
where $[\cdot,\cdot]$ indicates the commutator between quantum mechanical operators and $\alpha$ is a positive constant that removes the singularity of the linear system and forces the solution to be orthogonal to the valence manyfold, using standard techniques from density-functional perturbation theory (DFPT) \cite{RevModPhys.73.515}. In order to avoid alignment problems between  wave-functions at different time steps, $\dot\phi_v^\conduction$ is calculated by moving the derivative to the projector \cite{Marcolongo2014,QE2}, using the relation:
\begin{equation}
    \begin{aligned}
        \left | \dot\phi_v^\conduction \right> &\equiv \hat{P}^\conduction\left|\dot\phi_v\right> \\
        &= {\hat{P}}^\conduction  \dot{\hat{P}}^\valence\left|\phi_v\right>,
    \end{aligned}
\end{equation}
which can be evaluated using a finite-difference scheme, as explained in detail in Sec. \ref{sec:code_structure}, reading:
\begin{equation}
    \begin{aligned}
       \left | \dot\phi_v^\conduction \right> &= \hat{P}^\conduction\dot {\hat{P}}^\valence\left|\phi_v\right> \\
       & \sim \frac{1}{dt}(1-\hat{P}^\valence(t))(\hat{P}^\valence(t+dt/2)-\hat{P}^\valence(t-dt/2))
       \left|\phi_v(t)\right>,
     \end{aligned}
   \label{eq:dotphi_v^c}
\end{equation}
where $dt$ is a time-discretization parameter, input of a \qeheat computation, which is discussed in more detail in Section \ref{sec:code_structure} and \ref{sec:dt_ren_2p}. After evaluating Eq. \eqref{eq:dotphi_v^c} and solving Eq.~\eqref{eq:barphi_v^c}, the results can be inserted into~\eqref{eq:J_KS_exp}. We note that the computational cost to evaluate all the components of the MUB current is dominated by the solution of the linear system, Eq. \eqref{eq:barphi_v^c}.

\subsection{Zero current} \label{sec:Zero_current}

The \emph{Zero} current $\bm{J}^0$, \cref{eq:J_0}, can be better treated by separating the local and non-local contributions from the ionic pseudo-potential, $\hat v_0$. Each contribution acts on a generic wave-function $\phi$ in the following way:
\begin{align}
\hat v^0 &=\sum_{s\bm L}  \hat v_{s\bm L}^{LOC}  + \hat v_{s\bm L}^{NL}, \\
\langle \bm r | \hat v_{s\bm L}^{LOC} |\phi\rangle &= f^{LOC}_s\left(|\bm{r}-\bm R_{s}- \bm L|\right) \phi (\bm r), \label{eq:localpp1}\\
\hat v_{s\bm L}^{NL} |\phi \rangle &=\sum_{lm} D^{s}_{l}
| \beta_{lm}^{s\bm L} \rangle \langle \beta_{lm}^{s\bm L} | \phi \rangle,
\label{eq:nonlocalpp}
\end{align}
where the total pseudopotential has been separated into atomic contributions and into its local (LOC), long-tailed,  and non-local (NL), short-range, components. The subscripts in the expressions $\hat v_{s\bm L}^{LOC}$ and $\hat v_{s\bm L}^{NL}$ indicate that the corresponding ionic pseudo-potential is centered at the atomic position $\bm R_s + \bm L$. Note that the atomic contributions $\hat v_{s\bm L}^{LOC/NL}$, when differentiated with respect to ionic positions, only depend on the position of the atom located at $\bm R_s+\bm L$. 

The local and nonlocal potentials provide two contributions to the Zero current, which we discuss individually in the next sections. For the local contribution, $f^{LOC}_s(r)$ is the radial local pseudo-potential provided in the pseudo-potential datasets for each species. The centered beta functions $\beta^s_{lm}(\bm r)$ (also called projectors) define the non-local component of the pseudo-potentials and are defined as $\beta^s_{lm}(\bm r) \equiv \beta^s_{l}(r)Y_{lm}(\hat {\bm r})$, where $Y_{lm}$ are the real spherical harmonics with quantum numbers $l,m$ and $\hat {\bm r}$ is the unit versor, not to be confused with the multiplicative position operator. We denote the radial components of the beta functions with the similar notation $\beta^s_{l}(r)$. These are the ones provided for each atomic species in the pseudo-potential datasets, alongside the constant $D$ matrix. At every time step, the beta functions need to be centered on the instantaneous ionic positions and in Eq. \eqref{eq:nonlocalpp} we indicated the translated beta function for atom at position ${\bm R_{s}-\bm L}$ with an apex. More explicitly :
\begin{equation}
    \langle \bm r|\beta_{lm}^{s \bm L} \rangle \equiv \beta^s_{lm}(\bm r-\bm R_{s} - \bm L).    
\end{equation}
We fix some handy notation and define from a (real) localized function $\gamma(\bm  r)$ (\emph{e.g.} a $\beta$ function or the local pseudo-potential) its periodic counterpart as:
\begin{equation}
    \overline \gamma \bm  (\bm  r) \equiv \sum_{\bm  L}\gamma(\bm  r - \bm  L), \label{eq:gammabar}
\end{equation}
whose Fourier components can be computed as:
\begin{align}
    \tilde{\overline \gamma} (\bm  G)= 
    \mathcal{F}[\gamma](\bm  G)  \equiv \frac{1}{\Omega} \int_{\mathbb{R}^3} \gamma(\bm  r)e^{-i\bm G\cdot\bm r} d\bm  r,
    \label{eq:periodic-counterpart}
\end{align}
where we introduced the symbol  $\mathcal{F}[\gamma](\bm  G)$ to identify a standard Fourier transform over all the three dimensional space, here defined for localized functions. In the following we use the notation $\overline \gamma^{\bm a}$ to indicate the translated and periodic function built from its localized counterpart, analogously to the notation introduced for the projector. Note that the following relation is used in the code, $\tilde{\overline \gamma}^{\bm a}(\bm G)= e^{-i\bm G\cdot\bm a} \mathcal{F}[\gamma](\bm  G)$, thanks to standard properties of the Fourier transform. The remaining Fourier transforms are  evaluated numerically by \texttt{QEHeat} whenever needed, as explained below.

\subsubsection{Zero current: local contribution} \label{sec:Zero_current-local}
The local pseudo-potential of the $s$-atom $f^{LOC}_s(r)$ behaves as $\sim -Z_s e^2/r$ for large $r$. We call this local long-range contribution to the flux $\bm{J}^{LR}_{0}$:
\begin{equation}
\bm{J}^{LR}_{0}=\sum_v \left \langle\phi_v \Bigl |  \sum_{s\bm L} (\bm{\hat r} - \bm  R_{s} - \bm L) \left(\bm  V_s \cdot \nabla_{s \bm L } f_s^{LOC}(|\bm{\hat r}-\bm R_{s} - \bm L|) \right) \Bigr | \phi_v \right \rangle. \label{eq:jzero_loc}
\end{equation}
We use the chain rule and the definitions:
\begin{align}
    h^s_{ij}(\bm  r) &\equiv \sum_{\bm  L}\frac{(\bm  r -\bm  L)_i(\bm  r -\bm  L)_j}{|\bm  r -\bm  L|}f^{'LOC}_{s}(|\bm  r-\bm  L|) \\ u_i(\bm  r) &\equiv -\sum_s \sum_{j\in \{x,y,z\}} V_{sj} h^s_{ij}(\bm  r-\bm  R_s),
\end{align}
where $f^{'LOC}_s(r)$ is the derivative of the local pseudo-potential. Note that both $h$ and $u$ are periodic functions. The current can then be rewritten as:
\begin{align}
    J_{0,i}^{LR} &=\int_{\Omega} n(\bm  r)u_i(\bm  r) d \bm  r \nonumber \\ &= \Omega \sum_{\bm  G} \tilde{n} (\bm  G) \tilde{u}_i(-\bm  G), \label{divquattro}
\end{align}
where:
\begin{align}
    \tilde{u}_i(\boldsymbol{G})=-\sum_{s} \sum_{j\in \{x,y,z\}} V_{sj} \tilde{h}_{ij}^{s}(\boldsymbol{G}) e^{-i \bm G \cdot \bm{R}_{s}}.
\end{align}

The reciprocal Fourier components of $h^s_{ij}(\bm  r)$ are computed through the following procedure, which avoids an explicit numerical differentiation of the pseudo-potential. One writes:
\begin{align}
    h^s_{ij}(\bm  r) &= \partial_j\left[\sum_{\bm  L} (\bm  r-\bm  L)_i f^{LOC}_s(|\bm  r-\bm  L|)\right]-\delta_{ij}\sum_{\bm  L} f^{LOC}_s(|\bm  r-\bm  L|) \nonumber \\
    &\equiv \sum_{\bm G } \left( \tilde{h}^{1,s}_{ij}(\bm  G)+ \tilde{h}^{2,s}_{ij}(\bm  G)  \right) e^{i\bm G\cdot\bm r}.
\end{align}
The resulting expressions of $\tilde{h}^1(\bm G)$ and $\tilde{h}^2(\bm G)$ can be evaluted using Eq. \eqref{eq:periodic-counterpart} and the standard expansion of $e^{-i\bm G\cdot\bm r}$ into spherical harmonics and Bessel functions $J_l$:
\begin{equation}
   \tilde{h}^{1,s}_{ij}(\bm  G) =
    \begin{dcases}
        0 & \text{for } \bm  G=\bm  0 \\
        \frac{4\pi}{\Omega}\frac{G_i G_j}{G^2}G  \int_0^{\infty} r^3 f^{LOC}_s(r)J_1(Gr)dr
        & \text{for } \bm  G \ne \bm  0,
    \end{dcases}
\end{equation}
where the $\bm  G=0$ component vanishes thanks to the presence of a derivative in the definition of $h^1$. For $h^2$ the situation is different:
\begin{equation}
    \tilde{h}^{2,s}_{ij}(\bm  G) =
    \begin{dcases}
      -\delta_{ij}\frac{4\pi}{\Omega}
\int_0^{\infty} r^2 f^{LOC}_s(r) dr
& \text{for } \bm  G=\bm  0 \\
        -\delta_{ij}\frac{4\pi}{\Omega}
\int_0^{\infty} r^2 f^{LOC}_s(r) J_0(Gr)dr
        & \text{for } \bm  G \ne \bm  0
    \end{dcases}
\end{equation}
These expressions need still to be a bit elaborated before being evaluated by \texttt{QEHeat}. In order to evaluate integrals of the localized functions, one needs to add and subtract the asymptotic long-range tail of the local pseudo-potential. The long-range part can be integrated analytically for any finite value of the Yukawa screening parameter, $\mu$, after plugging in the exact form of the spherical Bessel functions. For $h2$, this permits to extract the divergent part in the $\bm  G=\bm  0$ contribution as well. The final results read:
\begin{equation}
    \tilde{h}^{1,s}_{ij}(\bm  G) = \\
    \begin{cases}
        0 & \text{for } \bm  G=\bm  0 \\[5pt]
        \displaystyle \frac{4\pi}{\Omega}\frac{G_i G_j}{G^2} G \left[ \int_0^{\infty} r^3 \left(f^{LOC}_s(r)+\frac{e^2Z_s}{r}\right)J_1(Gr)dr
            - \frac{2e^2Z_s}{G^3} \right]
        & \text{for } \bm  G \ne \bm  0
    \end{cases}
\end{equation}
and
\begin{equation}
    \tilde{h}^{2,s}_{ij}(\bm  G) = \\
    \begin{cases}
        \displaystyle -\delta_{ij}\frac{4\pi}{\Omega}\left[\int_{0}^{\infty}r^2\left( f^{LOC}_s(r)+\frac{e^2Z_s}{r} \right) dr -\frac{e^2Z_s }{\mu^2} \right] & \text{for } \bm  G=\bm  0 \\[15pt]
        \displaystyle -\delta_{ij}\frac{4\pi}{\Omega}
\left[\int_0^{\infty} r^2\left( f^{LOC}_s(r)+\frac{e^2Z_s}{r} \right)J_0(Gr)dr -\frac{e^2Z_s}{G^2}\right]
        & \text{for } \bm  G \ne \bm  0
    \end{cases}
\end{equation}
We note here again that only from $h2$ we get a Coulombian divergence when $\mu \rightarrow 0$. Overall, the divergent part of the Zero current is equal to
\begin{equation}\label{eq:J0div}
    \bm J_{div}^0=-Z_{t o t}e^2 \frac{4 \pi}{\mu^2 \Omega} \sum_{s} \bm V_{s} Z_{s},
\end{equation}
where $Z_{tot}=\sum_s Z_s$.

\subsubsection{Zero current: nonlocal contribution}
The non-local part is inherently short range and we call it $\bm{J}^{SR}_{0}$:
\begin{equation}
\bm{J}^{SR}_{0}=\sum_{v}  \left\langle\phi_v \left | \sum_{s\bm L} (\bm{\hat r} - \bm  R_{s} - \bm L) \left(\bm  V_s \cdot \nabla_{s \bm L } \hat v_{s\bm L}^{NL}\right) \right | \phi_v \right \rangle, \label{eq:jzero_nonloc}
\end{equation}
For a pair of two localized functions $(\gamma_1(\bm r),\gamma_2(\bm r))$ we introduce the notation, given a generic translation $\bm a$:
\begin{align}
&\mathcal{A}[\gamma_1,\gamma_2](\bm  a) \equiv \sum_{v} \langle \overline \gamma_1^{\bm a} | \phi_v \rangle \langle\phi_v | \overline \gamma_2^{\bm a}\rangle = \langle \overline \gamma_1^{\bm a} | \hat P_{v}|\overline \gamma_2^{\bm a}\rangle,
\label{eq:block}
\end{align}
where we recall that $\overline \gamma_1^{\bm a}$ and  $\overline \gamma_2^{\bm a}$ are the translated and periodic counterpart of localized functions. 
We note that, once the Fourier components of $\phi$, $\overline \gamma_1$ and  $\overline \gamma_2$ are known, evaluating $\mathcal{A} [\gamma_1 , \gamma_2](\bm  a)$ involves just scalar products between periodic functions and is straightforward. Using these expressions and expanding the projector operator in their integral form, the current can be rewritten as:
\begin{align}
J^{SR}_{0,i}=\sum_{s} \sum_{
lm} \sum_{j\in \{x,y,z\}}   V_{s j}D^s_{l}
\left( \mathcal{A}[-r_i\partial_j\beta^s_{lm} , \beta^s_{lm}](\bm  R_s)+  \mathcal{A}[r_i\beta^s_{lm} , -\partial_j\beta^s_{lm}](\bm  R_s)\right). \label{eq:nonloca-final}
\end{align}
In order to evaluate Eq. \eqref{eq:nonloca-final} via \eqref{eq:block} the Fourier transforms of the following four localized functions are needed : $\beta_{lm}(\boldsymbol{r}), r_{i} \beta_{lm}(\boldsymbol{r}) , -\partial_{i} \beta_{lm}(\boldsymbol{r}) , -r_{i} \partial_{j} \beta_{lm}(\boldsymbol{r}) $, where we dropped the atomic index $s$. We need only to evaluate the first two expressions thanks to the identities:
\begin{align}
\mathcal{F}\left[-\partial_{j} \gamma\right](\bm G)& =-i G_{j}\mathcal{F}\left[\gamma\right](\bm G) \\
\mathcal{F} \left[-r_{i} \partial_{j} \gamma \right](\bm G) &=-i G_{j}\mathcal{F}\left[r_i \gamma\right](\bm G)+\delta_{i,j}\mathcal{F}\left[\gamma\right](\bm G),
\end{align}
which are valid for every localized function $\gamma(\bm r)$. We show how to evaluate the Fourier transform of the second function,  $r_{i} \beta_{lm}(\boldsymbol{r})$,  for $i=x$, which is the most complex one. The procedure is similar for the remaining expressions. One starts by replacing the factor $x$  using that $Y_{11}\left(\hat {\bm r}\right)=-\sqrt{\frac{3}{4 \pi}} \frac{x}{r}$  (according to the convention followed by \qe\ for the sign of the spherical harmonics). By expanding $e^{-i\bm G\cdot\bm r}$ as well into spherical harmonics, one gets:
\begin{multline}
\mathcal{F}[x\beta_{lm}](\boldsymbol{G})=-\frac{4 \pi}{\Omega} \sqrt{\frac{4 \pi}{3}} \sum_{l^{\prime} m^{\prime}} Y_{l^{\prime} m^{\prime}}\left(\hat{\bm G}\right) \times \\
\left((-i)^{l^{\prime}} \int_{0}^{\infty} r^{3} \beta_l(r) J_{l^{\prime}}(G r) d r\right)\left(\int d \hat{\bm r} \ Y_{l^{\prime} m^{\prime}}\left(\hat{\bm r}\right) Y_{lm}\left(\hat{\bm r}\right) Y_{11}\left(\hat{\bm r}\right)\right), \label{eq:nonlocal-fourier}
\end{multline}
where $d \hat{\bm r}$ indicates an integral over the solid angle, such that $d \bm r=r^2 d \hat{\bm r} dr$. In \qe\ the Clebsch-Gordan coefficients are stored in a array \texttt{ap} defined as:
\begin{equation}
    Y_{l m} Y_{l' m'}=\sum_{LM} \mathtt{ap}\left(L, M, l, m, l', m'\right) Y_{LM}
\end{equation}
Application of this relation to the product $Y_{lm}Y_{11}$ allows us to simplify Eq. \eqref{eq:nonlocal-fourier} to the final form:
\begin{equation}
    \mathcal{F}[x\beta_{lm}](\boldsymbol{G})=-\frac{4 \pi}{\Omega} \sqrt{\frac{4 \pi}{3}} \sum_{LM} Y_{LM}\left(\hat{\bm G}\right) \left((-i)^{L} \int_{0}^{\infty} r^{3} \beta_l(r) J_{L}(G r) d r\right) \mathtt{ap}\left(L,M, l, m, 1, 1 \right).
\end{equation}
The remaining radial integral is performed numerically on a grid. 
\subsection{Ionic current} \label{sec:ionic_current}
The so-called Ionic flux, $\bm{J}^n$, is the contribution to the total energy flux that depends only on the ionic positions, $\bm{R}_s$, and velocities, $\bm{V}_s$. First of all let's take, as reported in \ref{sec:Notation}, the definition of the ionic energy $e_s^0 = \frac{1}{2}M_s V_s^2 + w_s$ and the electrostatic energy:
\begin{equation}\label{eq:wi_L}
    w_s = \frac{e^2}{2} \sum_{t\neq s}\sum_{\bm L}\frac{Z_tZ_s}{\vert \bm{R}_s-\bm{R}_t-\bm{L}\vert} + \frac{1}{2}e^2 Z_s^2\sum_{L\neq 0}\frac{1}{L},
\end{equation}
where $L=\vert \bm{L} \vert$. We can separate the expression of the Ionic flux, \cref{eq:J_n}, in two terms: one depending only on the mass and velocity of the ions, an other depending on $w_s$ and its gradient.
\begin{equation}\label{eq:w'}
   \nabla_{t \bm L }w_s = -\frac{e^2}{2}Z_s Z_t\frac{\bm{R}_s-\bm{R}_t-\bm{L}}{\vert \bm{R}_s-\bm{R}_t-\bm{L}\vert}f'(\vert \bm{R}_s-\bm{R}_t-\bm{L}\vert),
\end{equation}
where $\nabla_{t \bm L }$ is the shorthand notation for the gradient with respect to displacement of the atom with position $\bm R_t + \bm L$, introduced in \cref{sec:Theory}. We, also, introduced $f(x) = \frac{1}{x}$ to keep track of the Coulombian contributions when applying the Yukawa screening.

For the sake of simplicity, let's introduce the following four quantities:
\begin{align}
   \overline{S}^{A}_{ij} &=  \sum_{\bm L \neq \bm 0 }\frac{ L_{i} L_{j}}{\vert \bm{L}\vert}f'(L), \label{eq:Sa}\\
   S^{B} &=  \sum_{ \bm L \neq \bm 0}f(L) \label{eq:Sb}, \\
   S^C(\bm R_s-\bm R _t) &=  \sum_{\bm L}f(\vert \bm{R}_s-\bm{R}_t-\bm{L} \vert) \label{eq:Sc}, \\
   \overline{S}^D_{ij}(\bm R_s-\bm R _t) &=  \sum_{\bm L }\frac{(\bm R_s-\bm R _t-\bm L)_{i}(\bm R_s-\bm R _t-\bm L)_{j}}{\vert \bm{R}_s-\bm{R}_t-\bm{L}\vert}f'(\vert \bm{R}_s-\bm{R}_t-\bm{L}\vert), \label{eq:Sd}
\end{align}
where $i,j \in \{x,y,z\}$ represent the Cartesian coordinates. Then, consider the following properties for the function $f(x)$:
\begin{align}
   & \partial_i f(|\bm  x-\bm  L|)=\frac{(\bm  x-\bm  L)_i}{|\bm  x-\bm  L|}f'(|\bm  x-\bm  L|),
\end{align}
and that  $\overline{S}^{A}_{ij} = \lim_{x \to 0} \sum_{\bm L \neq \bm 0 }\frac{(\bm x-\bm L)_{i}(\bm x-\bm L)_{j}}{\vert \bm x-\bm{L}\vert}f'(\vert \bm x-\bm{L}\vert) $. The following relations between $\overline{S}^{A}_{ij}$, $S^B$, $S^C(\bm R_s-\bm R _t)$ and $\overline{S}^D_{ij}(\bm R_s-\bm R _t)$ hold:
\begin{gather}
    \begin{aligned}
        \overline{S}^{A}_{ij} &= \lim_{x\to 0}\left[  \partial_{ x_j}\sum_{\bm L\neq \bm 0 }(\bm x-\bm L)_{i}f(\vert \bm x-\bm{L}\vert)\right ] - S^{B} \delta_{ij} \\
        & \equiv S^A_{ij}-S^{B}\delta_{ij},
    \end{aligned}\label{eq:Sa_bar} \\
    \begin{aligned}
        \overline{S}^D_{ij}(\bm R_s-\bm R _t) & = \left [  \partial_{\bm x_j}\sum_{\bm L }(\bm x-\bm L)_{i}f(\vert \bm x-\bm{L}\vert) \right]_{x=\bm R_s-\bm R _t} - S^{C}(\bm R_s-\bm R _t) \delta_{ij} \\
        & \equiv S^D_{ij}(\bm R_s-\bm R _t)-S^{C}(\bm R_s-\bm R _t)\delta_{ij},
    \end{aligned}\label{eq:Sd_bar}
\end{gather}
Finally, by combining \cref{eq:wi_L,eq:w',eq:Sa,eq:Sb,eq:Sc,eq:Sd,eq:Sa_bar,eq:Sd_bar} into \cref{eq:J_n} we can re-write $\bm{J}^n$:
\begin{align}
       & \bm{J}^{n} = \bm J^{nA} + \bm J^{nB} + \bm J^{nC} + \bm J^{nD} \label{eq:Jall} \\
       & \bm J^{nA} = \frac{1}{2}\sum_{s} \bm{V}_s  M_s V_s^2 \label{eq:J_a^ion}\\
       &  J_{i}^{nB} = \sum_{s} V_{s i} Z_s^2 e^2 S^B -\frac{e^2}{2} \sum_{s} \sum_{j \in x,y,z} V_{s j} Z_s^2 S^A_{ij} \label{eq:J_b^ion}\\
       & \bm J^{nC} = \sum_s\sum_{t \neq s}  Z_t  Z_s e^2 S^C(\bm R_s-\bm R _t) \bm{V}_s \label{eq:J_c^ion}\\
       &  J^{nD}_{i} =- \frac{e^2}{2}\sum_{s}\sum_{t \neq s}Z_t Z_s \sum_{j \in x,y,z} S^D_{ij}( \bm R_s-\bm R _t) V_{tj} \label{eq:J_d^ion}.
\end{align}
Introducing the matrix $T_{ij}=S^B \delta_{ij} + S^A_{ij}$, which only depends on the cell and not on the individual atoms within it, we can rewrite \cref{eq:J_b^ion} as a sum of terms that depends only on the atomic species and the cell:
\begin{align}
     J_{i}^{nB} =\frac{e^2}{2} \sum_{j\in \{x,y,z\} }T_{ij} \sum_{S} Z_S^2 \sum_{t\in S} V_{t j} \label{eq:J_b^ion_tmass},
\end{align}
where $\sum_{S}$ is a sum over the atomic species. Recent theoretical developments in the statistical analysis of fluxes \cite{Marcolongo2016,Baroni2020,Ercole2016,Marcolongo2020,Bertossa2019}, and in particular the so called \emph{gauge} \cite{Marcolongo2016} and \emph{convective} \cite{Bertossa2019} \emph{invariances} tell us that fluxes written as sum of terms that depends only on the atomic species do not contribute to the value of the thermal transport coefficient. Thus, $\bm  J^{nB}$ can be neglected in the computation of $\bm{J}^n$, reducing the implemented formula to:
\begin{equation}
    \bm{J}^{n} = \bm J^{nA} + \bm J^{nC} + \bm J^{nD}.
\end{equation}

We still need to deal with the long-range Coulomb contributions in $  \bm J^{nC}$, $\bm J^{nD}$ and $\bm J^{nB}$. Following the scheme explained in \cref{sec:Theory} we introduce a Yukawa screened interaction, $\frac{1}{x} \rightarrow \frac{e^{-\mu x}}{x}$, and we will study the limit $\mu \rightarrow 0$. We can, then, straightforwardly apply the Ewald techniques \cite{grosso2000solid}, with a convergence parameter $\eta$, for the computation of $S^A_{ij}$, $S^B$, $S^C(\bm R_s-\bm R _t)$ and $S^D_{ij}(\bm R_s-\bm R _t)$. Further details of the computation can be found in \ref{sec:App_reciprocal}:
\begin{align}
    S^A_{ij}  = &  - \sum_{\bm  L \neq \bm 0} \frac{L_i L_j}{L^2} \left[ f(L) \mathrm{erfc}(\sqrt{\eta}L) +
2\sqrt{\frac{\eta}{\pi}}e^{-\eta L^2} \right]-2\delta_{ij}\sqrt{\frac{\eta}{\pi}}
\notag\\ 
 & + \delta_{ij} \sum_{ \bm L \neq  \bm 0} f(L) \mathrm{erfc}(\sqrt{\eta}L) +
   \sum_{\bm  G \ne \bm  0}\frac{4\pi}{\Omega}\frac{G_i G_j}{G^2}\frac{\exp(\frac{-G^2}{4\eta})}{G^2}\left[2+\frac{G^2}{2\eta}\right], \label{eq:Sa_G} \\
S^B &= \sum_{ \bm L \neq  \bm 0} f(L) \erfc(\sqrt{\eta}L) - 2 \sqrt{\frac{\eta}{\pi}} +\frac{4 \pi}{\Omega} \sum_{ \bm G \ne \bm 0} \frac{e^{-\frac{G^2}{4\eta}}}{G^2} +
\frac{4\pi}{\Omega} \left(\frac{1}{\mu^2} - \frac{1}{4\eta}\right), \label{eq:Sb_G} \\
S^C(\bm R_s-\bm R _t) &=  \sum_{\bm  L} f(|\bm  R_s-\bm R _t-\bm  L|)\erfc(\sqrt{\eta}|\bm  R_s-\bm R _t-\bm  L|) + \nonumber \\ & \qquad\qquad\qquad \qquad\qquad \frac{4 \pi}{\Omega} \sum_{\bm  G \ne \bm 0} \frac{e^{\frac{-G^2}{4\eta}}}{G^2} e^{i\bm  G  (\bm R_s-\bm R _t)} + \frac{4\pi}{\Omega} \left(\frac{1}{\mu^2}-\frac{1}{4\eta}\right),  \label{eq:J_c^ion_G}\\
S^D_{ij} (\bm R_s-\bm R _t) &=   \sum_{\bm  L}\left( \sqrt{\eta}h(\sqrt{\eta}|\bm  R_s-\bm R _t-\bm  L|)\delta_{ij}+ \right. \notag \\
&   \quad\quad \left. \eta h'(\sqrt{\eta}|\bm  R_s-\bm R _t-\bm  L|)\frac{(\bm  R_s-\bm R _t-\bm  L)_i (\bm  R_s-\bm R _t-\bm  L)_j }{|\bm  R_s-\bm R _t-\bm  L|} \right)+   \notag \\
 &  \qquad\qquad \qquad\qquad  \qquad +\sum_{\bm  G \ne \bm  0} e^{i\bm  G \bm ( \bm R_s-\bm R _t)} \frac{4\pi}{\Omega}\frac{G_i G_j}{G^2}\frac{e^{-\frac{G^2}{4\eta}}}{G^2}\left(2+\frac{G^2}{2\eta}\right), \label{eq:Sd_G}
\end{align}
where $\erfc(x)=1-\erf(x)$, $\erf(x)$ is the error function \cite{grosso2000solid}, and $h(x)=\frac{\erfc(x)}{x}$. 
Since \cref{eq:Sa_G,eq:Sd_G} do not diverge in $\mu$, the divergent parts of the Ionic flux are given only by \cref{eq:Sb_G,eq:J_c^ion_G}:
\begin{align}
 \bm J_{div}^{nB} &= e^2\frac{4 \pi}{\Omega \mu^2}\sum_s \bm V_s Z_s^2, \label{eq:Jdiv_b} \\
 \bm{J}_{div}^{nC}  &= e^2 \frac{4\pi }{\mu^2\Omega}\sum_{s} \sum_{t \neq s } Z_s Z_t  \bm{V}_s, \label{eq:Jdiv_c} \\
 \bm{J}_{div}^n  &= \bm J_{div}^{nB} +\bm{J}_{div}^{nC},  \nonumber \\& =  Z_{t o t} e^2 \frac{4\pi }{\mu^2\Omega}\sum_{s} Z_s  \bm{V}_s. \label{eq:J_ntot_div} 
\end{align}


\subsection{Hartree and Exchange-correlation currents}
A finite-difference scheme, explained in detail in section \ref{sec:code_structure}, can be directly implemented to evaluate the scalar fields $\dot v^H(\bm r)$ and $\dot n(\bm r)$, which are needed to evaluate $\bm J^{H}$ and $\bm J^{XC}$. For the Hartree current, the gradient $\bm \nabla v_H(\bm r)$ is needed as well. Since the gradient operator in reciprocal space is multiplicative, this suggests to rewrite the entire expression in reciprocal space:
\begin{equation}
    \bm J^H = -i\frac{\Omega}{4\pi e^2}  \sum_{\bm G} \dot v^H(\bm G)  v^H(-\bm G) \bm G,
\end{equation}
which is the equation actually implemented. Analogously, for $\bm J^{XC}$ the gradient $\bm \nabla n(\bm r)$ is first computed in reciprocal space and then Fourier transformed onto the real grid. In the PBE case, $\epsilon_{GGA}$ has an explicit analytical expression as a function of $n$ and $|\bm \nabla n|$. The analytic expression of the derivative $\partial \epsilon_{GGA}/\partial(\bm \nabla n)$ is cumbersome but can be straightforwardly derived from the latter. The resulting expression is then evaluated at the local values of $n(\bm r_i)$ and $\bm \nabla n(\bm r_s)$ for each point $\bm r_s$ of the grid in real space. Finally, contributions from all grid points are summed up.

\subsection{Electronic density current}
As a by-product of the computation of the adiabatic energy flux, \qeheat also evaluates the adiabatic electron-number flux by implementing Thouless' expression \cite{thouless83} and using DFPT \cite{BaroniDFPT}. To this end, by leveraging the continuity equation, one first formally expresses the number flux as the first moment of the time derivative of the electron number density, to obtain for every Cartesian component $i$:
\begin{align}
    \bm{J}^{el}_i & =\int\dot n(\bm{r})r_id\bm{r} \nonumber \\
    & =\sum_v \left<\phi_v|{\hat r}_i|\dot\phi_v \right>+\left<\dot\phi_v|{\hat r}_i|\phi_v \right> \nonumber \\
    &=2\sum_v \left<\barphi\,_i|\dot\phi_v^\conduction\right>.
    \label{eq:el-current}
\end{align}
All quantities  needed to evaluate the Electronic density current have already been discussed in the section dedicated to the Kohn-Sham current. The electronic flux thus evaluated is interesting \emph{per se}, \emph{e.g.} to compute the electric conductivity in ionic conductors, and also as an ingredient to facilitate the statistical analysis of the energy-flux time series, using \emph{multi-component} \cite{Bertossa2019}  or \emph{decorrelation} \cite{Marcolongo2020} techniques.

\subsection{Center-of-mass ionic current and Charge current}
The code outputs a trivial but useful current defined for each atomic species as
\begin{equation}
    \bm J^{CM}_S=\sum_{t\in S}\bm V_t,
\end{equation}
where $S$ is the atomic species index. The sum is over all atoms of kind $S$.
This current can be used both for data analysis or for computing the Charge current together with Eq.~\eqref{eq:el-current}:
\begin{equation}
    \bm J^Q=-e\bm J^{el}+\sum_{t}\bm v_teZ_t=-e\bm J^{el}+e\sum_{S}Z_S\bm J^{CM}_S
\end{equation}
where $eZ_S$ is the pseudo-potential charge of the atom of species $S$, and $\bm J^{el}$ is the Electronic density current computed by the code, defined in Eq.~\eqref{eq:el-current}.

\subsection{Divergences}\label{sec:Divergences}
In this section, we discuss the divergences arisen in the computations of Ewald sums in \cref{sec:Zero_current-local,sec:ionic_current}. First of all, we note that $\bm{J}_{div}^0+\bm{J}_{div}^n=0$, showing that the expression for the total MUB flux is free of any divergent term, as we already stated in \cref{sec:Theory}.
We highlight that any divergent term, being $\bm{J}_{div}^0$, $\bm{J}_{div}^{nB}$ or $\bm{J}_{div}^{nC}$, it can be written as a sum of terms depending only on the atomic species, precisely as $\bm{J}_{div}^{nB}$. Thus, invoking the same \emph{invariance principles} that allowed us to neglect $\bm{J}_{div}^{nB}$, we can state that any divergent contribution is non diffusive and would not contribute to the transport coefficient.

\section{Reciprocal space computation of the $S^{A}_{ij}$ and $S^B$ } \label{sec:App_reciprocal}

In the following appendix we show explicitly the computation of $S^A_{ij}$, $S^B$. $S^C$ and $S^D_{ij}$ can be computed applying more straightforwardly standard Ewald techniques \cite{grosso2000solid}, then the full derivation is left to reader.

Recalling the definition of $S^A_{ij}$ from \cref{eq:Sa_bar}:

\begin{equation}
    \begin{aligned}
        S^A_{ij} &= \lim_{x\to 0}  \partial_{{\bm x}_{j}} \sum_{\bm L \neq \bm 0}(\bm x-\bm L)_{i}f(\vert \bm x-\bm{L}\vert) \\
        & = \lim_{x\to 0} \partial_{{\bm x}_{j}}\sum_{\bm L \neq \bm 0}(\bm x-\bm L)_{i}f(\vert \bm x-\bm{L}\vert)\Bigl ( \mathrm{erf}(\sqrt{\eta} \vert \bm x-\bm{L}) + \mathrm{erfc}(\sqrt{\eta} \vert \bm x-\bm{L}) \Bigr )  .
    \end{aligned}
\end{equation}

The term containing $\mathrm{erfc}(\sqrt{\eta} \vert \bm x-\bm{L}\vert )$ can be simply computed in real space: 
\begin{equation}\label{eq:lim_spazio_diretto}
\begin{aligned}
    \lim_{x\to 0}
    \partial_{{\bm x}_{j}} \sum_{\bm L \neq \bm 0}(\bm x-\bm L)_{i}f(\vert \bm x-\bm{L}\vert)\mathrm{erfc}(\sqrt{\eta}  \vert \bm x-\bm{L}\vert )
    = -\sum_{\bm L \neq \bm 0} \frac{L_{i}L_{j}}{L^2}  & \left( \frac{\mathrm{erfc}(\sqrt{\eta}L)}{L} +2\sqrt{\frac{\eta}{\pi}} e^{- \eta L^2}\right)+ \\
    & +\delta_{ij} \sum_{\bm L \neq \bm 0} \frac{\mathrm{erfc}(\sqrt{\eta}L)}{L} ,
    \end{aligned}
\end{equation}
while:

\begin{multline}\label{eq:limite}
    \lim_{x\to 0} \partial_{{\bm x}_{j}}\sum_{\bm L \neq \bm 0}(\bm x-\bm L)_{i}f(\vert \bm x-\bm{L}\vert)\mathrm{erf}(\sqrt{\eta}  \vert \bm x-\bm{L}\vert ) = \\
    \lim_{x\to 0} \partial_{{\bm x}_{j}}\sum_{\bm L }(\bm x-\bm L)_{i}f(\vert \bm x-\bm{L}\vert)\mathrm{erf}(\sqrt{\eta}  \vert \bm x-\bm{L}\vert ) - 
    \lim_{x\to 0} \partial_{{\bm x}_{j}}  \left(  x_{i} \mathrm{erf}(\sqrt{\eta }x) f(x) \right).
\end{multline}

The first term of \cref{eq:limite} can be easily computed in reciprocal space, while it can be demonstrated that $\lim_{x\to 0}  \partial_{\bm x_{j}}\left(x_{i} erf(\sqrt{\eta} x) f(x) \right)= 2 \delta_{ij}\sqrt{\frac{\eta}{\pi}}$.
Summing all these pieces together we get the expression in \cref{eq:Sa_G}:

\begin{align}
     S^A_{ij}  = &  - \sum_{\bm  L \neq \bm 0} \frac{L_i L_j}{L^2} \left[ \frac{\mathrm{erfc}(\sqrt{\eta}L)}{L}+
2\sqrt{\frac{\eta}{\pi}}e^{-\eta L^2} \right]-2\delta_{ij}\sqrt{\frac{\eta}{\pi}}
\notag\\ 
 & + \delta_{ij} \sum_{\bm  L \neq \bm 0} \frac{\mathrm{erfc}(\sqrt{\eta}L)}{L} +
   \sum_{\bm  G \ne \bm  0}\frac{4\pi}{\Omega}\frac{G_i G_j}{G^2}\frac{\exp(\frac{-G^2}{4\eta})}{G^2}\left[2+\frac{G^2}{2\eta}\right].
\end{align}

$S^B$ is, instead, defined as:
\begin{equation}
    S^{B} =  \sum_{ \bm L\neq \bm 0}f(L),
\end{equation}
introducing a functional dependence on $x$:

\begin{equation}
    \begin{aligned}
        S^{B} & =  \lim_{x\to 0} \sum_{\bm L\neq \bm 0}f(\vert \bm x-\bm{L}\vert) = \\
        & = \lim_{x\to 0} \sum_{ \bm L\neq \bm 0}f(\vert \bm x-\bm{L}\vert) \mathrm{erf}(\sqrt\eta \vert \bm x-\bm{L}\vert ) +\lim_{x\to 0} \sum_{\bm L\neq \bm 0}f(\vert \bm x-\bm{L}\vert) \mathrm{erfc}(\sqrt\eta \vert \bm x-\bm{L}\vert ).
    \end{aligned}
\end{equation}
The second expression can be computed in direct space as it is, while the first requires some further work:
\begin{align}\label{eq:Sb_limi_erf}
    \lim_{x\to 0} \sum_{\bm L\neq \bm 0}f(\vert \bm x-\bm{L}\vert) \mathrm{erf}(\sqrt{\eta} \vert \bm x-\bm{L}\vert )
    &= \lim_{x\to 0} \sum_{\bm L}f(\vert \bm x-\bm{L}\vert) \mathrm{erf}(\sqrt{\eta} \vert \bm x-\bm{L}\vert ) - \lim_{x\to 0} f(x) \mathrm{erf}(\sqrt{\eta}x) \nonumber \\
    &= \lim_{x\to 0} \sum_{\bm L}f(\vert \bm x-\bm{L}\vert) \mathrm{erf}(\sqrt{\eta} \vert \bm x-\bm{L}\vert ) - 2\sqrt{\frac{\eta}{\pi}}.
\end{align}
Now the first term of \cref{eq:Sb_limi_erf} can be computed in reciprocal space, thus, summing all the contributions, we can regain the expression in \cref{eq:Sb_G} for $S^B$:
\begin{equation}
    S^B =    \sum_{ \bm L \neq \bm 0} f(L)\mathrm{erfc}(\sqrt{\eta}L)-2\sqrt{\frac{\eta}{\pi}} +\frac{4 \pi}{\Omega} \sum_{  \bm G \ne \bm 0} \frac{\exp(\frac{-G^2}{4\eta})}{G^2} +
\frac{4\pi}{\Omega}\left(\frac{1}{\mu^2}-\frac{1}{4\eta}\right).
\end{equation}
Finally we remark that the two terms computed in this Appendix are general and work for generic cells. However, since $\bm J^{nB}$ does not contribute to the value of thermal transport coefficient, up to the present version of \qeheat\ we only implemented the simpler expression for cubic cells:

\begin{equation}
    \begin{aligned}
        \bm J^{nB} = & \sum_{s} \bm V_{s} Z_s^2 e^2 ( S^B -\frac{1}{2} S^A) = \\ 
        & \sum_{s} \bm V_{s} Z_s^2 e^2 \left( \frac{2}{3} \sum_{\bm L\neq \bm 0} f(L)erfc(\sqrt{\eta}L) +\frac{8\pi}{3\Omega}\sum_{ \bm G \ne  \bm 0} \frac{\exp(\frac{-G^2}{4\eta})}{G^2} - \frac{4}{3}\sqrt{\frac{\eta}{\pi}}-\frac{2\pi}{3\eta \Omega}+\frac{4\pi}{\Omega \mu^2}\right),
    \end{aligned}
\end{equation}
where we defined:
\begin{align}
S^A \equiv & \frac{1}{3} \mathrm{Tr}[S^A_{ij}],
\end{align}
$\mathrm{Tr}[\cdot]$ indicating the trace of a matrix.
The following identity can be used to recover the formula for cubic systems from the general one:
\begin{align}
\sum_{ \bm L }e^{-\eta L^2} =& \frac{\sqrt{\pi^3}}{\Omega \sqrt{\eta}^3}\sum_{ \bm  G }\exp\left(\frac{-G^2}{4\eta}\right) 
\end{align}
\section{Implementation check of individual currents}

\begin{figure}
    \centering
    \includegraphics[width=0.8\textwidth]{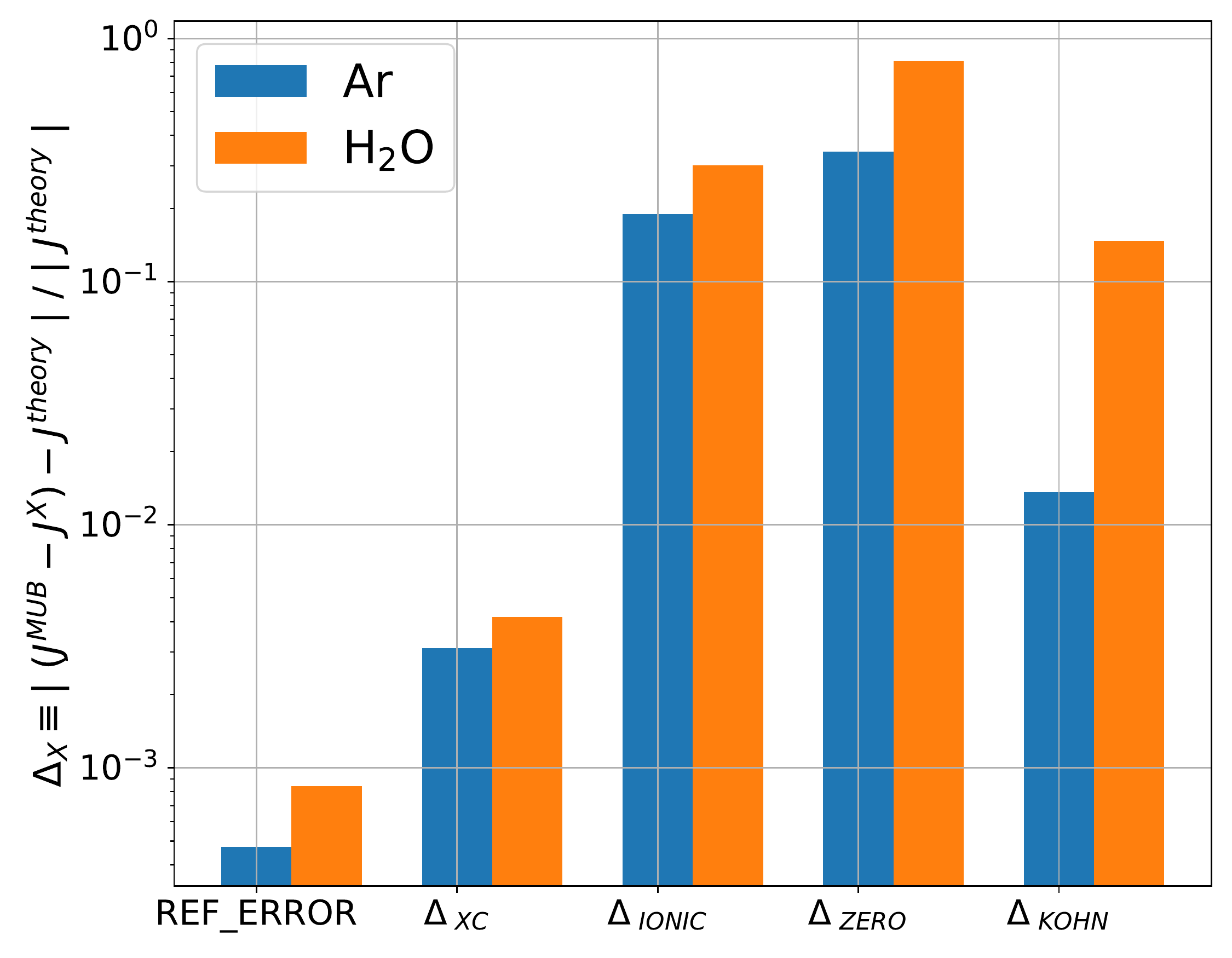}
    \caption{In the case of a finite system at equilibrium translating at constant speed $\bm v$, the current $\bm J^{MUB}$ must be equal to  $\bm J^{theory} \equiv E^{tot} \times \bm v$. We report in the picture on the leftmost histogram  $\text{REF}\_\text{ERROR} = | \bm J^{MUB} - \bm J^{theory} | / |\bm J^{theory}|$ in the case of an Argon atom and a water molecule translating at constant speed. Ideally $\text{REF}\_\text{ERROR}$ should be identically zero, but for numerical reasons it can be just a small value. To check that  $\text{REF}\_\text{ERROR}$ is indeed small and at the same time to validate the correct implementation of each individual current, $\Delta_{X}$ represents the same quantity after the substitution, in the numerator of the fraction, $\bm J^{MUB} \rightarrow \bm J^{MUB} - \bm J^{X}, X \in \{\text{XC},\text{IONIC},\text{ZERO},\text{KOHN}\}$. Since removing a current component increases the error w.r.t. the expected value significantly, this proves the correct implementation of each individual current. See text for the parameters used in the computations.}
    \label{fig:translation}
\end{figure}

As discussed in the text, in the case of a finite system at equilibrium translating at constant speed $\bm v$, the current $\bm J^{MUB}$ must be equal to  $E^{tot} \times \bm v$. In Fig. \ref{fig:translation} we report the modulus of $\bm J^{MUB}-E^{tot}\bm v$, normalized by the module of $E^{tot}\bm v$ and indicated with $\text{REF}\_\text{ERROR}$. In order to 
check the correct implementation of each individual current, we also report values of $\Delta_{X}$ , which represent the same quantity after the substitution $\bm J^{MUB} \rightarrow \bm J^{MUB} - \bm J^{X}, X \in   \{\text{XC},\text{IONIC},\text{ZERO},\text{KOHN}\}$. If the current $X$ is correctly computed, the error with respect to the reference value $E^{tot}\bm v$ should increase, as is indeed observed. For Argon, we used the QE parameters $\text{ecut}=160 \ Ry$ and $\text{econv}=10^{-16} \ Ry$. For water $\text{ecut}=120 Ry$, $\text{econv}=10^{-14} Ry$. We used a cubic simulation cell of $20$ and $30 \AA$ for Argon and water respectively. 

\section{Numerical stability of \qeheat}\label{sec:dt_ren_2p}

In the following appendix we show the stability and convergence properties of a \qeheat\ calculation on a snapshot of 125 water molecules. 

\begin{figure}[bth]
    \centering
   \includegraphics[width=0.8\textwidth]{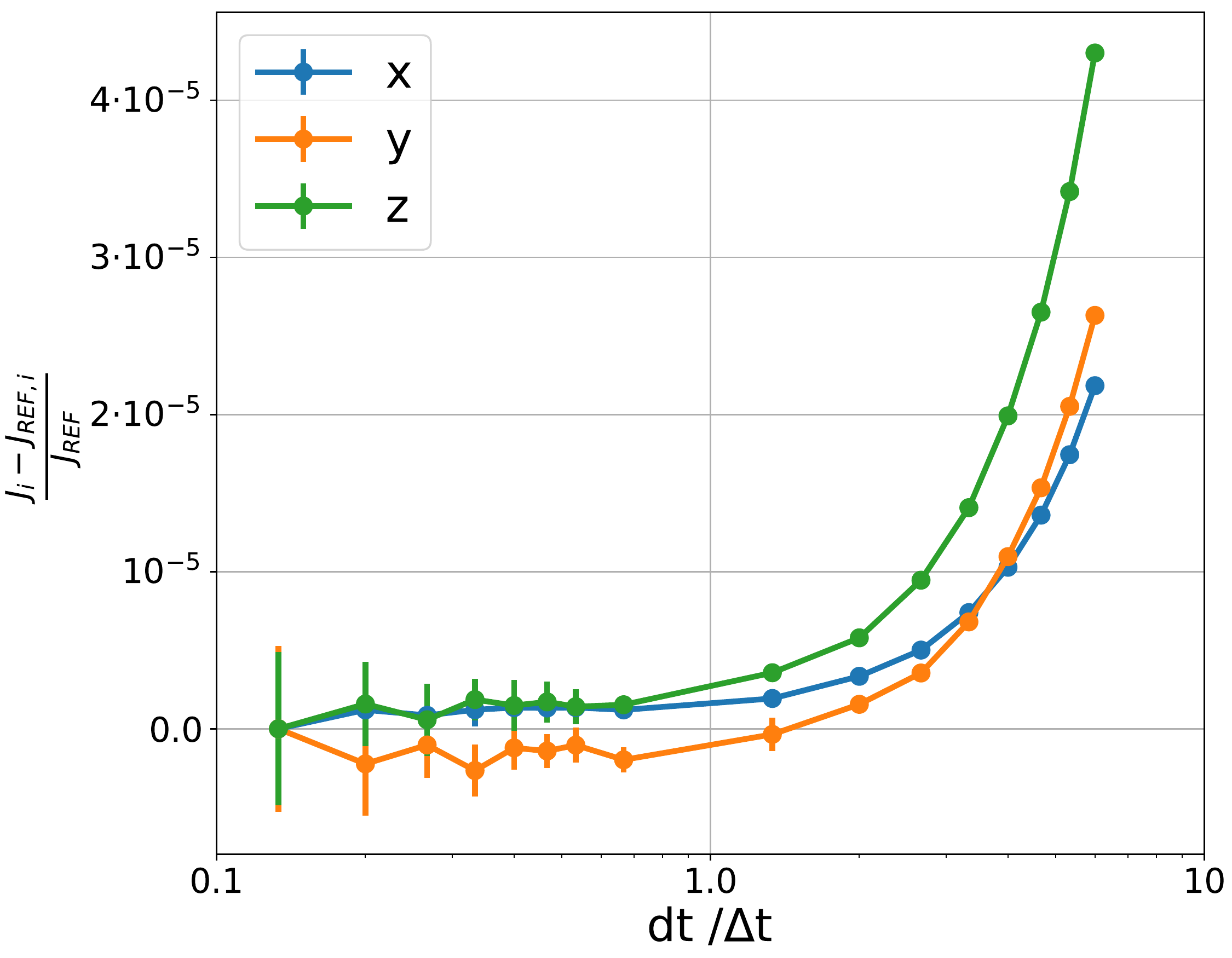}
    \caption{The behaviour of $( J_{i}(dt)- J_{\text{REF},i} )/ J_{\text{REF}}$ as a function of the timestep $dt$ used in the symmetric derivative for a snapshot of H$_2$O, where 
    $J_{\text{REF}} = |{\bm J}_{\text{REF}}| $
    and $i\in \{x,y,z \}$. The symbol $J_{i}(dt)$ refers to a component of the MUB flux computed with a discretization step $dt$, as reported in the $x$-axis. The reference value $J_{\text{REF},i}$ is evaluated for each coordinate at the smallest value of $dt$ available, in the example taken to be with $dt/\Delta t=0.13$. The $x$-axis is in units of the simulation step $\Delta t$. The picture shows that the code can handle correctly even small values of $dt$. At the same time, it shows that the non-linear effects due to high values of $dt$ are negligible up to the range of $dt$ explored, and in particular negligible at $dt=2\Delta t$. The errorbar are an estimate of the statistical uncertainty computed using the testing feature presented in \cref{sec:UsageBenchmarks} and using 20 different initialization of the wavefunctions.} 
\label{fig:3p_econv}
\end{figure}

\cref{fig:3p_econv} reports a scaled version of the three Cartesian components of the MUB energy flux $J_{i}(dt),i \in \{x,z,y\}$, as a function of the time-discretization parameter $dt$. The $x-$axis is in units of the Car-Parinello MD simulation timestep, indicated with $\Delta t$. For each Cartesian coordinate, the behaviour of the error $( J_{i}(dt)- J_{\text{REF},i} )/ J_{\text{REF}}$, considering a reference and scale values, is reported, thus showing possible non-linear contributions due to the choice of a large $dt$. 
The reference value $J_{\text{REF},i}$ is evaluated for each coordinate at the smallest value of $dt$ available and $J_{REF} = |\bm J_{\text{REF}}|$.

\cref{fig:3p_econv} shows the presence of small non-linear effects for higher values of $dt$, and that, at least for the presented system, nonlinear effects do not take off substantially up to $dt=2\Delta t$. At this value of $dt$, even reducing it by a factor of 10 would change the component of the current by less than 0.001\%, a negligible error given that a typical value of the thermal transport coefficient has an accuracy of 10\% \cite{Marcolongo2014,Bertossa2019,Grasselli2020}. Moreover $dt=2\Delta t$ would be beneficial in an on-the-fly computation, allowing to reuse the same wavefunctions computed in the MD simulation, neglecting the need for the recomputation of the scf cycles. The errorbars in the figure are computed using the testing feature, provided with \qeheat,  presented in \cref{sec:UsageBenchmarks} and averaging 20 fluxes obtained from different initial random wave-functions.

\begin{figure}[htb]
    \centering
    \includegraphics[width=0.8\textwidth]{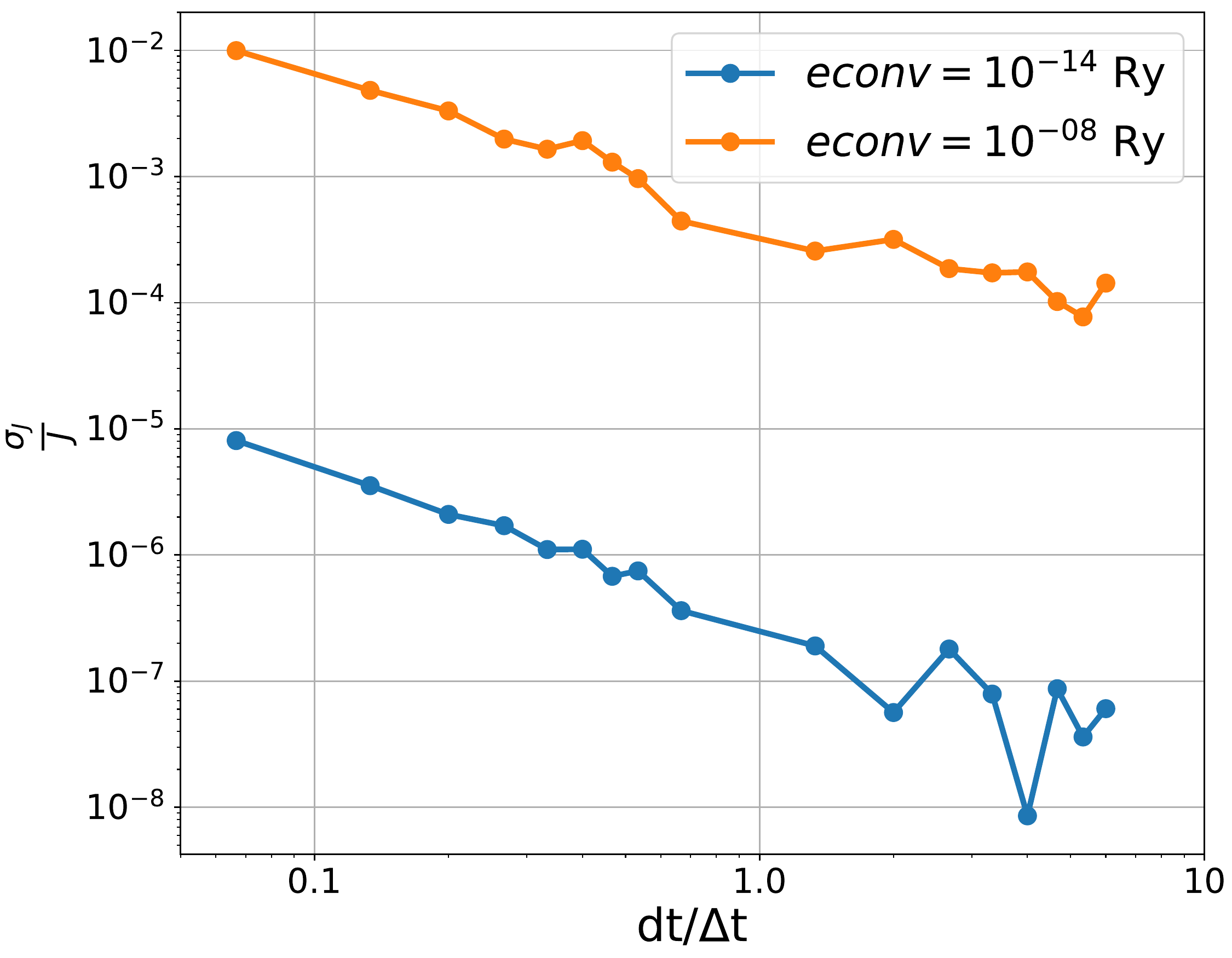}
    \caption{Percentage error on the x component of the total energy flux for a selected snapshot of H$_2$O . The source of error here considered is due to the inherent noise on the wave-functions computed by \qe\ 
    and was estimated restarting \qeheat\ $20-$times with different seeds. The error is plotted for different values of \texttt{econv} as function of the timestep $dt$ used in the symmetric derivative. Note that the error decreases decreasing \texttt{econv}, as expected, but also increasing $dt$. Not shown here is the onset of the non-linear behavior due to a too large $dt$, visible, instead, in \cref{fig:3p_econv}. (see also \ref{sec:dt_ren_2p}). Different colors represent different values of \texttt{econv}.}
    \label{fig:3p_err}
\end{figure}

The previous calculations were performed with the self-consistent threshold \texttt{econv} equal to  $10^{-14}~$Ry. The lower, the better the quality of the wavefunctions calculated.  In order to further test and prove the stability of the MUB current,  we used again the aforementioned testing feature to compute the $dt$ dependence of the statistical uncertainty of the MUB current,  at different values of \texttt{econv}. \cref{fig:3p_err} shows the percentage error of the $x$-component of the flux for a specific snapshot of 125 molecule. Even though the dependence on $dt$ is similar for the two values of \texttt{econv}, it is clear that increasing \texttt{econv}, at a fixed $dt$, increases the statistical error by orders of magnitude, three orders when \texttt{econv} is changed from $10^{-14}~$Ry to $10^{-08}~$Ry and at any fixed $dt$. Note also that the variance decreases when increasing $dt$, in an exponential way. It must be stated that the errors in \cref{fig:3p_err} do not include the effect of the non-linearity due to a too large $dt$, bringing a (small) bias to the estimation of the MUB flux which can be seen, for example, in \cref{fig:3p_econv}.

\begin{figure}[hbt]
    \centering
    \includegraphics[width=0.8\textwidth]{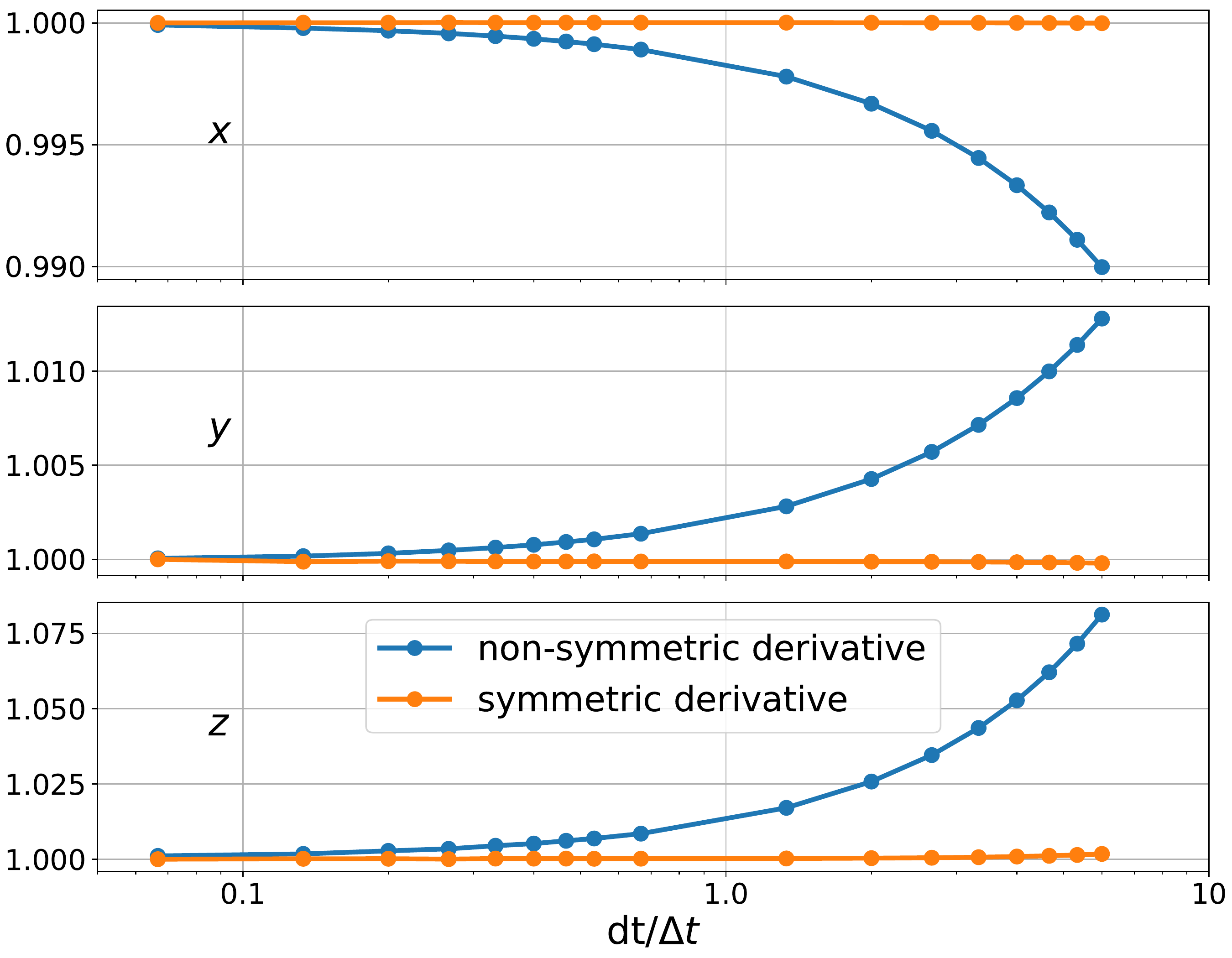}
    \caption{The behaviour of $J_{i}(dt)/ J_{\text{REF}}$ as function of the timestep $dt$ used in the numerical derivative. The three panels compare the results obtained with the non and symmetric derivative, each panel shows a different component of the energy current $J_i$. $J_{\text{REF}}$, taken as reference, is the modulus of MUB flux computed for $dt=0.66\Delta t$ and the symmetric derivative approach. The errorbars are obtained repeating the computation of the current with a different initial wave-function. The picture also shows that for the non-symmetric derivative there are only slightly stronger non-linear behavior on the current for large $dt$. }
    \label{fig:2p_vs_3p_der}
\end{figure}

In \cref{sec:code_structure} we showed how \qeheat\ implements numerical derivatives with the symmetric approach. The code allows also, simply changing the \verb.three_point_derivative. keyword to \verb|false|, to compute the derivative within a non-symmetric approach. In this case the implemented expression is:
\begin{equation}
    \dot f(\{ \bm  R_s\}) \approx
        \frac{f(\{ \bm  R_s\})-f ( \{\bm R_s-\bm  V_sdt\})}{dt},
\end{equation}

 removing one scf computation with respect to the symmetric derivative scheme, thus reducing slightly the computational time. Quantities that are not numerically time-derived are evaluated with atoms at positions $\{\bf R_s$\}. For the same snapshot of water of \cref{fig:3p_err} and $ecut=85~$ Ry, $econv=10^{-14}~$Ry, we computed the energy current with both the non and symmetric derivative approaches. \Cref{fig:2p_vs_3p_der} shows that the latter returns an energy flux by far more numerically stable with $dt$. However, in both cases, the values of the current  only slowly deviate from a constant behaviour, after increasing dt. Even the 2-point derivative gives results that differ of few percentage points.

\section{Computational Cost}\label{sec:Comp_Cost}

The computational cost of a QEHeat calculation depends on several factors that one should take into consideration, e.g.:
\begin{itemize}
    \item \emph{Typical decaying times.} Decaying times, defined as time lengths when the relevant autocorrelation functions become negligible, can vary a lot according to the system. Longer decaying times require longer simulations to acquire enough statistics. Typically, liquid materials show smaller decaying times. Other convenient situations are solids at high pressures and temperatures, or glasses. See also Fig. \ref{fig:Water_IceX_vs_ps} in main text.
    \item \emph{Sampling frequency}. It is not necessary to compute the energy flux for every single step of the molecular dynamics trajectory,  in order to avoid unneeded wastes of computer time. The optimal sampling frequency can depend on the chosen methodology to extract the thermal conductivity coefficient from the time series of the energy flux. Cepstral theory helps in deciding an optimal sampling frequency ( see e.g. the analysis reported in Supp materials ). In the example for liquid water of Sec. \ref{sec:calculationExample}, we evaluated the heat flux every 60 units of time in Hartree atomic unit, i.e. 20 Car-Parrinello molecular dynamics steps in our simulation.
    \item \emph{Simulation cell sizes.}
    One has to take into account that strongly harmonic systems, like crystalline solids at ambient temperature, may require large simulation cells to remove boundary effects. For high temperature or high pressure solids this should be a smaller issue. Anharhmonic effects in disordered systems or glasses can also reduce the typical path lengths and reduce therefore boundary effects.
    \item \emph{Required accuracy.} All settings should be tuned according to the desired accuracy on the thermal conductivity coefficient, which may vary according to the application.
\end{itemize}
We provide here the overall computational cost of the computation presented in section \ref{sec:calculationExample} for liquid water, as a guideline, even if system specific setups are suggested. For this test, we chose conservative parameters and the default symmetric derivative scheme. We ran the $MD$ calculations with $4$ nodes, $192$ processors, whereas the \qeheat\ calculations, being trivially parallelizable, where each run on $1$ node and $48$ processors. All times here reported are multiplied by the number of processors and identify therefore the total cost of the computation.\\
For the 64-molecules system, a general MD step costs $\approx 1.2 \times 10^2$s of cpu time (average over $\approx 2.5 \times 10^5$ steps, leading to a trajectory $240$ps long).  
The cost of evaluating $\bm J^{MUB}$ for a single snapshot was $\approx 1.5 \times 10^3$s of cpu time, in which $84\%$ of the time is reserved to the solution of the linear system in \cref{eq:J_KS} during the evaluation of $\bm J^{KS}$. The $3$ minimizations needed for the symmetric derivative scheme cost all together $\approx 2.2 \times 10^3$s (mean over all PW calculations). 
 We evaluated $\bm J^{MUB}$ every $20$ MD timesteps, thus the overall overhead was around $1.5$ times the cost of the whole \textit{ab initio} molecular dynamics simulation. \\ 
To conclude, at the present state our code recomputes the self-consistent cycles for each step, adding some extra time that we considered in this analysis. We are currently working on using directly the wafefunctions provided during the Car-Parrinello molecular dynamics (CP-MD) simulation, a promising feature considering that our analysis on the $dt$ dependence implies that the discretization step can be chosen equal to the CP-MD time step, and it will be available in a future release. All the computations are done on the Tier-0 system called Marconi CINECA, which have 3188 nodes equipped with 2 Intel Xeon 8160 (SkyLake) at 2.10 GHz with 24-cores each \cite{MARCONI}.
\\

\end{document}